\documentclass[epj]{svjour}
\usepackage{latexsym}
\usepackage{graphics}
\usepackage{amssymb}
\usepackage{amsfonts}
\usepackage{mathrsfs}
\usepackage{bm}
\usepackage{amsmath}
\usepackage{color}
\usepackage[dvipsnames]{xcolor}
\usepackage{arydshln} 
\usepackage{booktabs}
\usepackage{multirow}
\usepackage[title]{appendix}

\begin{document}

\title{An approximation to the Woods-Saxon potential based on a contact interaction}

\author{C. Romaniega\inst{1}, M. Gadella\inst{1}, R.M.   Id Betan\inst{2,3,4} \and L.M. Nieto\inst{1}}

\institute{Departamento de F\'{\i}sica Te\'{o}rica, At\'{o}mica y \'{O}ptica and IMUVA, Universidad de Valladolid, 47011. Valladolid, Spain. \and  Instituto de F\'isica Rosario (CONICET-UNR), Bv. 27 de Febrero 210 bis, S2000EZP Rosario. Argentina. \and  Facultad de Ciencias Exactas, Ingenier\'ia y Agrimensura (UNR), Av. Pellegrini 250, S2000BTP Rosario. Argentina. \and Instituto de Estudios Nucleares y Radiaciones Ionizantes (UNR), Riobamba y Berutti, S2000EKA Rosario. Argentina}

\date{Received: date / Revised version: date}

\abstract{
We study a non-relativistic particle subject to a three-dimensional spherical potential consisting of a  finite  well and a radial $\delta$-$\delta'$ contact interaction  at the well edge.  
This contact potential is defined by appropriate matching conditions for the radial functions, thereby fixing a self adjoint extension of the non-singular Hamiltonian. 
Since this model admits exact solutions for the wave function, we are able to characterize and calculate the number of  bound states. We also extend some well-known properties of certain spherically symmetric potentials and describe the resonances, defined as unstable quantum states. Based on the Woods-Saxon potential, this configuration is implemented as a first approximation for a mean-field nuclear model. The results derived are tested with experimental and numerical data  in the double magic nuclei $^{132}$Sn and  $^{208}$Pb with an extra neutron. 
 }

\authorrunning{Romaniega, Gadella, Id Betan \& Nieto}

\titlerunning{An approximation to Woods-Saxon potential based on a contact interaction} 

\maketitle

\section{Introduction}
\label{intro}

In one-dimensional non-relativistic quantum mechanics, point potentials or potentials supported on one or a discrete collection of points, like the Dirac delta interactions, have deserved considerable attention recently (see  \cite{BR} and references quoted therein). These potentials are often exactly solvable and therefore, provide a good insight for some quantum phenomena like scattering. In addition, they serve as a fair approximation for various types of interactions, as very short range interactions between a single particle and a fixed heavy source as well as a contact interaction in the centre of mass of two particles. This is the origin of the name {\it contact potentials}. They also function as suitable approximations when the particle  wavelength is much larger than the range of the potential. In spite of their simplicity, they have a vast amount of applications in modelling real physical systems, as we can see for instance in a recent review \cite{BR}, some books \cite{DO,ALB} and the references therein. The landmark example in solid state physics is a limiting case of the Kronig-Penney model in which a countably infinite set of Dirac delta interactions are periodically distributed along a straight line \cite{KRP,KIT,ERMAN}. 
Other examples of physical interest are the following: a Bose-Einstein condensation in a harmonic trap with a tight and deep ``dimple'' potential, modelled by a Dirac delta function \cite{Uncu}; a non-perturbative study of the entanglement of two directed polymers subject to repulsive interactions given by a Dirac delta  potential \cite{Ferrari}; light propagation in a  one-dimensional realistic  dielectric superlattice, modelled by means of a periodic array of these functions  for the cases of transverse electric, transverse magnetic, and omnidirectional polarization modes \cite{Lin}.

From a purely mathematical point of view, one-dimen\-sion\-al contact potentials have been studied as self adjoint extensions of the kinetic energy operator $-d^2/dx^2$ \cite{ALB,KU}.  This approach has been used to construct several one-dimensional models  which go beyond the Dirac delta potential \cite{G}. A  discussion on the physical meaning of the one-dimensional contact potentials constructed as self adjoint extensions of the kinetic energy operator is given in \cite{KP}. It is remarkable that, inspired in the physics of contact interactions, new mathematics has been developed \cite{AK}.

In the present manuscript we consider a three-dimen\-sion\-al spherically symmetric potential. Although the problem reduces to a one-dimensional one once the centrifugal term is included, the model can be more suitable for the implementation into realistic situations than the above-mentioned one-dimensional results. As will be explained later, this potential would serve as an approximation for a nuclear model.

Focusing on the contact interaction, and for reasons to be exposed below, in this paper we are going to analyse   a linear combination of two independent interactions: a Dirac delta and a $\delta'$ potential, both supported on a hollow sphere of radius $x_0$. Thus, this interaction is represented by a potential of the form

\begin{equation}\label{1.1}
V(x)=v_1\,\delta(x-x_0)+v_2\,\delta'(x-x_0), \quad x\geq 0,\ x_0>0.
\end{equation}
As mentioned earlier, spherically symmetric three-dimen\-sion\-al Schr\"odinger equations admit a one-dimensional counterpart if the orbital angular momentum is left fixed, which is the so called radial equation.
 When working with the radial equation, the   $\delta$-$\delta'$ interaction supported on the  sphere  become a  one-dimensional $\delta$-$\delta'$  interaction supported on the point $x_0>0$.
A self adjoint determination for the Hamiltonian $H=-d^2/dx^2+V(x)$ with the potential \eqref{1.1} in one dimension has been discussed in \cite{GNN}. Although the definition for the $\delta$ interaction is universally accepted, this is not the case for the  $\delta'$ term, for which at least two definitions  are consistent with its desirable properties \cite{AFR,FGGN}. The determination and sense of the $\delta'$ interaction is given by the use of proper matching conditions for the wave functions at $x_0$. These matching conditions determine a domain in which the Hamiltonian with the $\delta'$ interaction is self-adjoint \cite{KU}.  In this paper,  following the lines developed in \cite{GNN,GGN}, we shall use the so called {\it local} $\delta'$ interaction since it is compatible with the $\delta$ potential in such a way that the total Hamiltonian $H=-d^2/dx^2+V(x)$  is self adjoint. As we shall see, since both interactions are  supported on the same point, the proposed model naturally leads to this determination which is beautifully given by a proper choice of matching conditions for the wave functions at $x_0$. 

Accordingly, the same choice of the Hamiltonian $H=-d^2/dx^2+V(x)$ has been used in a study in which the interaction \eqref{1.1} plays a fundamental role: the approximation of a system formed by two thin plates in order to describe the quantum vacuum fluctuations in the presence of boundaries: the Casimir effect \cite{SRJUANITO}. In this context, the addition of the $\delta'$ term can be useful when dealing with boundary conditions (BC). Robin  BC  can be obtained as a finite limit of the previous interaction. Although Dirichlet  BC can  also be reached with the $\delta$ potential alone, the strength going to infinity could be troublesome  for the Casimir self-energy of a sphere \cite{graham2004dirichlet,barton2004casimir}.   Some other discussions on properties of contact potentials are given in \cite{AFR,Z,Z1,CAL} and their use in supersymmetric quantum mechanics is shown in \cite{ROS,CHIS,ERIK,ASDRUBAL1,ASDRUBAL2} and references quoted therein. We do not intend to be exhaustive and just mention recent literature.

As pointed out above, we also investigate its possible use as a mean field nuclear model, which has been a motivation to study this particular case. Within the Woods-Saxon approximation, the Dirac delta interaction has been  used for the calculation of resonant parameters and energy spectrum in \cite{2017delaMadrid}. Recently, it has also been employed to study the spectral function of the unbound nucleus $^{25}$O \cite{2018IdBetan}. In the latter case, a comparison of the spectrum   obtained between the $\delta$ potential versus the nuclear non-singular mean-field is performed. In this paper we also test the results obtained with the regular mean-field potential in two nuclei,  $^{133}$Sn and $^{209}$Pb.
 The main advantage of this approach is that the wave function can be easily solved in terms of well-known special functions. This enables us to derive analytic properties of the neutron energy levels structure.

The article is organized as follows. In Section 2, we introduce the potential under  consideration, which is written in a language that will fit well for the future application to a nuclear model. The radial $\delta$ and $\delta'$ terms appear here explicitly. 
In Section \ref{sec.se}, we derive the secular equation for which the solutions give the bound states. 
In Section \ref{section4}, we study the existence and localization of bound states, giving some rigorous results. We study the resonances arisen in our  example in Section \ref{section5}. The analysis of resonances is necessary because most of the known quantum states in nuclear or atomic physics are unstable. These findings are tested with the nuclei $^{133}$Sn and $^{209}$Pb in section \ref{sec.isotopes}, the latter being of some relevance in particle astrophysics  \cite{2012Sharrad}.
Concluding remarks, and two appendices, one devoted to some comments on the self adjointness of the Hamiltonian and the other to the proofs of the main results of the text, give an end to this paper.

\section{Model and motivation} \label{sec.mf}

Along this section, we develop the quantum model under study. We have preferred to use a language that makes it suitable for possible applications, particularly in nuclear physics.  Nevertheless, a non-relativistic quantum particle in a spherical well with a contact interaction in the edge can be analysed within this framework. 
In order to make the link between the strength which appears in the radial $\delta$-$\delta'$ interaction and real physical parameters \cite{1982Vertse,plb104,Machleidt}, we start with the following three-dimensional single particle Hamiltonian in the centre of mass system,
\begin{equation}\label{hbold}
      H(\mathbf r) = - \frac{\hbar^2}{2\mu} \nabla^2_{\mathbf r} 
              + U_0(r) 
              + U_\text{so}(r) (\mathbf L\cdot \mathbf S) 
              + U_\text{q}(r) .
\end{equation}
The reduced mass  is denoted by $\mu$, being $U_0(r)$, $U_\text{so}(r)$, and $U_\text{q}(r)$, essentially, the Woods-Saxon potential \cite{WS}, its first and second derivative, respectively,
\begin{eqnarray}
    U_0(r) &=& -V_0 \, f(r) 
    				=  -V_0 \, \frac{1}{1+e^{(r-R)/a}} , \\ [1ex]
    U_\text{so}(r) &=& \frac{V_\text{so}}{\hbar^2} \, f'(r)  
                   =- \frac{V_\text{so}}{a \hbar^2}  \, 
                \frac{e^{(r-R)/a}}{\left(1+e^{(r-R)/a} \right)^2},  \label{eq.Vso} \\ [1ex]
    U_\text{q}(r) &=&   V_\text{q} \, f''(r)
                             = -   \frac{V_\text{q}}{a^2} \,
                  \frac{ e^{(r-R)/a}\, \left( 1- e^{(r-R)/a} \right)}
                  {\left( 1+e^{(r-R)/a} \right)^3} . 
\end{eqnarray}
Since we only study configurations with an extra neutron,  the Coulombic potential is not included in the Hamiltonian.
The strengths $V_0$ and $V_\text{so}$ of the Woods-Saxon potential and the spin-orbit, respectively, are positive defined in order to reproduce the experimental magic number \cite{1969Bohr}, while the sign of $V_\text{q}$ can be selected  to fit with the experimental data.  This second derivative  may be employed as a form factor in the transition operator for the quadrupolar electric transition E2 in the Interacting Boson Approximation model \cite{Woude}. The nuclear radius $R$ is parametrized  in terms of the nuclear mass $A=N+Z$  as $R=r_0 A^{1/3}$, being $r_0$  constant and $N,\ Z$ the number of neutrons and protons, respectively. The parameter $a$ gives the thickness of the surface of $f(r)$.  The nuclear shell model considers $N$ or $Z$ as a magic number, and optimized all these parameters in a way to reproduce, as well as possible, the low-lying energy levels of the nuclei with one extra neutron or proton \cite{2007Schwierz}. Typical values for the parameters are  $r_0=1.27$ fm, $a=0.7$~fm, $V_0=51 \pm 33 (N-Z)/{A}$ MeV, with $+\, (-)$ for proton (neutron) \cite{2007Suhonen}. 

Going back to the Hamiltonian \eqref{hbold}, we rewrite the kinetic operator   in terms of the orbital angular momentum $\mathbf L$ and the radial coordinates as
\begin{equation}
   - \frac{\hbar^2}{2\mu} \nabla^2_{\mathbf r} = 
            - \frac{\hbar^2}{2\mu}\left[
            \frac{1}{r^2}\, \frac{\partial}{\partial r} \left( r^2 \frac{\partial}{\partial r} \right)
                        - \frac{\mathbf L^2/\hbar^2}{r^2}\right] .
\end{equation}
The eigenfunctions of the corresponding three-dimensional stationary Schr\"odinger equation 
 are factored into a radial ${u_{n\ell j}(r)}/{r}$ and angular part $\mathcal{Y}_{\ell jm}(\theta,\phi)$. The latter fulfils
\begin{eqnarray}
\mathbf L^2 \mathcal{Y}_{\ell jm}(\theta,\phi) =
  					\hbar^2\, \ell (\ell +1) \mathcal{Y}_{\ell jm}(\theta,\phi) , \qquad 
  ( \mathbf L \cdot \mathbf S)\, \mathcal{Y}_{\ell jm} (\theta,\phi) =  \hbar^2 \, \xi_{\ell j} \, \mathcal{Y}_{\ell jm} (\theta,\phi).        
\end{eqnarray}
The function $\mathcal{Y}_{\ell jm} (\theta,\phi)$, a linear combination of spherical harmonics $Y_{\ell m} (\theta,\phi)$, 
is a  simultaneous eigenfunction of the operators  $\mathbf L^2$, $\mathbf S^2$, $\mathbf J^2=(\mathbf L+\mathbf S)^2$ and $J_z$ \cite{2007Suhonen}. We have also defined 
$\xi_{\ell j} =\frac{1}{2} (j(j+1)-\ell(\ell+1)-\frac{3}{4} )$, that is
\begin{equation}\label{xiellj}
\xi_{\ell j} =\left\{
\begin{array}{ccc}
\displaystyle\frac{\ell }{2} & \textnormal{for} & \displaystyle j=\ell +\frac{1}{2}, \\ [2ex]
\displaystyle -\frac{(\ell +1)}{2} & \textnormal{for} & \displaystyle j=\ell -\frac{1}{2},
\end{array}\right.
\end{equation}
where $\ell$ takes values in the non-negative integers $\mathbb{N}_0$. For $\ell=0$ the only possibility is $j=1/2$ so $\xi_{0 j}=0$.
Using the above relations,  the radial part fulfils  $H u_{n\ell_j}(r) = E_{n\ell_j} \, u_{n\ell_j}(r)$ where
\begin{equation} \label{eq.h2}
  H  = \frac{-\hbar^2}{2\mu}
                                \left[
                                  \frac{d^2}{dr^2} - \frac{\ell (\ell +1)}{r^2} 
                                \right]
            - V_0 f(r)
            + V_\text{so} \xi_{\ell j}  f'(r) 
            + V_\text{q} f''(r).
\end{equation}
Observe that the spin-orbit interaction is defined without the  $1/r$ factor as usual \cite{1949Mayer,1949Haxel}. This may be done since the nuclear spin-orbit does not have the same origin as the one in the atom, and yet it is   not well understood. In table~\ref{table.1overr}, we show that the change has not effect on the $s$ partial wave, as it should be, while it is more pronounced as the principal quantum number increases for $\ell=1$. Since the difference can be absorbed in the effective strength $V_\text{so}$, the $1/r$ term  is  omitted in our definition of the spin-orbit form factor.

\begin{table}
\centering
\caption{\label{table.1overr} \small Neutron energy levels (MeV) in the core of $^{208}$Pb using $ 1/r \,V_\text{so}\, \xi_{\ell j}\, f'(r)$ (second column) versus $V_\text{so}\, \xi_{\ell j}\, f'(r)$ (third column) as the radial form factor of the spin-orbit interaction, with $a=0.7$ fm, $r_0=1.27$ fm, $V_0=44.4$ MeV, and $V_\text{so}=16.5$~MeVfm.}
\label{tab:1}      
\begin{tabular}{lll}
\hline\noalign{\smallskip}
State ($n\ell_j$) & $E^{*}_{n\ell_j}\ $ & $E_{n\ell_j}\ $  \\
\noalign{\smallskip}\hline\noalign{\smallskip}
$0s_{1/2}$ &   -40.231    &  -40.231  \\
$0p_{3/2}$ &   -36.328  & -37.078  \\
 $0p_{1/2}$ &  -35.928 &   -34.901   \\
 $1s_{1/2}$  &   -29.622 &  -29.622 \\
 $1p_{3/2}$    &   -23.471  &   -25.029 \\
 $1p_{1/2}$  &   -22.695  &   -20.134 \\
 $2s_{1/2}$   &   -15.299 &   -15.299 \\
 $2p_{3/2}$   &   \, -8.355 &   -10.303 \\
 $2p_{1/2}$  &   \,  -7.413  &  \,  -3.370 \\
\noalign{\smallskip}\hline
\end{tabular}
\end{table}
Finally,  in order to reach the explicit form of the effective potential used in section \ref{sec.isotopes}, we take the limit $a\to 0^+$.  Note that
\begin{equation}\label{1.4}
   \lim_{a\to 0^+} U_0(r) =   
   			\left\{ \begin{array}{ccc}  
					-V_0 & {\rm if}  & {r< R}  \\[1ex] 
				-V_0/2 & {\rm if}  & {r= R}  \\[1ex] 
					      0 & {\rm if}  & r>R      \end{array}   
		      \right\} 
		      = V_0\,[\theta(r-R)-1],   
\end{equation}
where $\theta(x)$ is the Heaviside step function. 
The function $f(r)$ can be seen as a distribution on a certain space of test functions, such as the Schwartz space. Then, if $\psi(r)$ is an arbitrary  function of this space  we denote the action of the distribution $f(r)$ on  $\psi(r)$ by $\langle\psi(r)|f(r)\rangle=\int_0^\infty \psi^*(r)\,f(r)\,dr$. For the first derivative we obtain
\begin{eqnarray}
  \lim_{a\to 0^+} \langle \psi(r) | \frac{d}{dr}f(r) \rangle
	       = -\lim_{a\to 0^+} \langle \psi'(r) | f(r) \rangle 
	   \      = - \langle \psi'(r) | 1-\theta(r-R) \rangle 
		    = \langle \psi(r) | -\delta(r-R)\rangle .  \label{1.5}
\end{eqnarray}
This holds since the Dirac delta is the derivative of the Heaviside step function, from the point of view of  distributions. Consequently, 

\begin{equation} \label{1.6}
		 \lim_{a\to 0^+}  V_\text{so}\, \xi_{\ell j} \, f'(r) = -  V_\text{so}\, \xi_{\ell j} \,\delta(r-R)\,.
\end{equation} 
In the same way, we obtain the following expression for the second derivative: 
\begin{equation} \label{eq.vdp}
     \lim_{a\to 0^+}  U_\text{q}(r) = \lim_{a\to 0^+}  V_\text{q}\, f''(r) =  -V_\text{q}\, \delta'(r-R)\,.
\end{equation}
In view of these considerations, the Hamiltonian \eqref{eq.h2} turns into
\begin{eqnarray}
  H_{\text{sing}}= -\frac{\hbar^2}{2\mu}
                                \left[
                                  \frac{d^2}{dr^2} - \frac{\ell(\ell+1)}{r^2} 
                                \right]
            + V_0\,[\theta(r-R)-1] -  V_\text{so}\, \xi_{\ell j} \,\delta(r-R)
             - V_\text{q}\, \delta'(r-R),   
              \label{eq.sh}
\end{eqnarray}
where the  singular (contact) terms are already included. A comment on the $\delta'$ contribution in \eqref{eq.sh} is in order here.  As explained in appendix~\ref{A}, the $\delta$-$\delta'$ perturbation is  defined using the formalism of self adjoint extensions of symmetric (formally Hermitian) operators with equal deficiency indices. This gives two options for the $\delta'$ term. The former is a $\delta'$ which is often called the {\it non-local $\delta'$} \cite{AFR}. However, this choice is incompatible with the Dirac-$\delta$ \cite{FGGN}, so that we have to use the other choice, the {\it local $\delta'$ interaction}, which is defined by matching conditions established at the point supporting the interaction.  From a distributional point of view, this is a generalization of the usual definition of the derivative of the delta, which has to be  adapted to test functions with a discontinuity at $x_0$ and a discontinuity of their derivative at the same point \cite{KU}. In any case, the $\delta'$ perturbation is properly defined via  matching conditions at $x_0$, as we have already mentioned.

Thus, bearing in mind that $R\gg a$, we can consider the above simplified one-dimensional Hamiltonian \eqref{eq.sh} as a mean-field potential to describe neutron energy levels. One of the main advantages  is that the eigenvalue equation $H_{\text{sing}} \, u_\ell (r) = E_{n\ell_j}\, u_\ell (r)$ can be solved exactly for the wave function\footnote[5]{For simplicity, and when no confusion arises, we will use abbreviated notation such as $u_\ell\equiv u_{n\ell_j}$.}
  in terms of Bessel functions. Consequently, the main findings of the text are based on the properties of these functions.   

\section{Solutions of the singular Schr\"odinger equation} \label{sec.se}

In this section, we determine  the eigenfunctions of the  singular Hamiltonian \eqref{eq.sh},
\begin{eqnarray} 
	\Bigl[	-\frac{d^2 }{dr^2}
 		+ \frac{\ell(\ell+1)}{r^2}
 				 -	\frac{2 \mu E}{\hbar^2} 
				  + \frac{2 \mu V_0}{\hbar^2} ( \theta(r-R)-1) 	
				  	     	          + \alpha \, \delta(r-R)
				   {+} \beta \, \delta'(r-R)
				    \Bigr] u(r)=0,
				    \label{2.2}
\end{eqnarray}
where
\begin{equation} \label{2.3}
		\alpha = -\frac{2 \mu}{\hbar^2} V_\text{so}\, \xi_{\ell j}  ,  
		\qquad 
		\beta = -\frac{2 \mu}{\hbar^2} V_\text{q} \,.
\end{equation}

At this point, we should remark that we may look to the parameters $\alpha$ and $\beta$ as two independent coefficients, with no relation whatsoever with any future application to a nuclear model. In this sense, equation \eqref{2.2} can be considered for a quantum particle subject to a spherical well with a $\delta$-$\delta'$ interaction at the edge. 

The radial  Schr\"odinger equation is defined on the interval $0 \le r <\infty$. Due to the presence of the contact potential, we divide this semi-axis into two regions:    $0 \le r < R$  and  $R < r$. We shall obtain the wave function in each region and then apply suitable matching conditions at $r=R$, thus defining the singular part of the Hamiltonian. 

\subsection{Interior wave equation}\label{sec:inside}

In the study of the solutions in the first region $0 \le r < R$, we  consider energy values  $E>-V_0$.  Hence, if we perform the transformations
$$
x=\gamma r\, \ y_\ell(x)= u_\ell(r)\,,\ \gamma= \frac{\sqrt{2{\mu}(V_0+E)}}{\hbar}\,, \  x\in[0,\gamma R),
$$
then,  equation \eqref{2.2}  becomes a Riccati-Bessel differential equation:
\begin{equation}\label{2.5}
\frac{d^2\,y_\ell(x)}{dx^2} -\frac{\ell(\ell+1)}{x^2}\,y_\ell(x)+y_\ell(x)=0\,, \qquad \ell\in\mathbb{N}_0.
\end{equation}
For each particular value of the orbital angular momentum  $\ell$, the general solution  is given by
\begin{equation}\label{2.6}
y_{\ell}(x)=\sqrt x \left(A_\ell\,J_{\ell+\frac12}(x)+B_\ell\, Y_{\ell+\frac12}(x)\right)\,,
\end{equation}
where $J_{\ell+ 1/2}(x)$ and $Y_{\ell+1/2}(x)$ denote the Bessel functions of first and second kind, respectively, being $A_\ell$ and $B_\ell$  arbitrary constants. For small values of the positive variable $x$, the asymptotic forms of the aforementioned Bessel functions are given by
\begin{equation}\label{2.7}
J_{\ell+\frac12}(x) \sim \frac{\left( \frac x2  \right)^{\ell+\frac12}}{\Gamma\left(\ell+\frac32\right)}, 
\qquad 
Y_{\ell+\frac12}(x) \sim -\frac{\Gamma\left(\ell+\frac12\right)}{\pi} \left( \frac 2x  \right)^{\ell+\frac12}.
\end{equation}
Hence, if we are looking for square integrable solutions, we should impose $B_\ell=0$ for  $\ell\ne 0$, since $\sqrt x\,Y_{\ell+1/2}(x)$ behaves near zero as $x^{-\ell}$. For $\ell=0$, the radial Hamiltonian is not self adjoint, although it admits a one parameter family of self adjoint extensions \cite{AJS}. To fix one of them, we need to set boundary conditions at the origin for the functions $y_0(x)$ in the domain of the radial Hamiltonian. The simplest possibility is $y_0(0)=0$, which forces the choice $B_0=0$. In consequence, $B_\ell=0$  $\forall \ell\in\mathbb{N}_0$.  On the other hand, it is obvious after \eqref{2.7} that $\sqrt x\,J_{\ell+\frac12}(x)$ is zero at the origin and therefore square integrable on the finite interval considered. Consequently, the admissible solutions are just
\begin{equation}\label{2.8}
u_{\ell}(r)= A_\ell \,\sqrt{\gamma r}\, J_{\ell+\frac12}(\gamma r)\,, \quad r\in[0,R), \quad \ell\in\mathbb{N}_0.
\end{equation}

\subsection{Exterior wave equation}\label{sec:outside}

For values of $r$ such that $R<r$, we have to solve the Schr\"odinger equation \eqref{2.2} for $V_0=0$. As we are looking for bound states, we require  $E<0$. Then, we first proceed with the following changes:
\begin{equation}\label{2.9}
z=\kappa r\,,\  y_\ell(z)=u_\ell(r)\,,\  \kappa=\frac{\sqrt{2\mu |E|}}{\hbar}\,,\   z\in (\kappa R,\infty)\,,
\end{equation}
which transform \eqref{2.2} into the following differential equation:
\begin{equation}\label{2.10}
\frac{d^2 y_\ell(z)}{dz^2}-\frac{\ell(\ell+1)}{z^2}\,y_\ell(z)-y_\ell(z)=0\,, \qquad \ell\in\mathbb{N}_0.
\end{equation}
For any value of $\ell$, the general solution of \eqref{2.10} is given by
\begin{equation}\label{2.11}
y_{\ell}(z)=\sqrt z \left(C_\ell\, I_{\ell+\frac12}(z)+D_\ell\, K_{\ell+\frac12}(z)\right).
\end{equation}
Here, $I_{\ell+1/2}(z)$ and $K_{\ell+1/2}(z)$ are the modified Bessel functions of first and second kind, respectively, being $ C_\ell$ and $D_\ell$  arbitrary constants. Again, if we are looking for square integrable solutions, we need to know the asymptotic behaviours of these functions for large values of $z$, which are,
\begin{equation}\label{2.12}
I_{\ell+\frac12}(z) \sim \frac{e^z}{\sqrt{2\pi z}}\,,\qquad K_{\ell+\frac12}(z) \sim \sqrt{\frac{\pi}{2z}}\,e^{-z}\,.
\end{equation}
Accordingly, the solution \eqref{2.11} is square integrable if, and only if, $C_\ell=0$. In this way, the only possible contribution comes from the second term, so that
\begin{equation}\label{2.13}
u_{\ell}(r)= D_\ell\,\sqrt{\kappa r}\,K_{\ell+\frac12}(\kappa r),\quad r\in (R,\infty)\quad \ell\in\mathbb{N}_0 .
\end{equation}
Once we have obtained the interior and exterior solutions, we need to link both of them at the point $r=R$ in an appropriate way.

\subsection{Matching conditions}\label{sec:matching}

As established by the standard bibliography on the subject \cite{KU}, there are requirements for the reduced radial function at the point $r=R$ which  fix a self adjoint determination of the operator  
\begin{equation*}
	-\frac{d^2}{dr^2}
 			 	+ \frac{\ell(\ell+1)}{r^2}  
				  + \frac{2 \mu V_0}{\hbar^2} [ \theta(r-R)-1] 
				     \,,
\end{equation*}
thus defining the final Hamiltonian of equation \eqref{2.2}. These requirements are given by matching conditions  relating the  function $u_{\ell}(r)$ and its first derivative at the limit values of $R$. They can be written in terms of a $SL(2, \mathbb{R})$ matrix as \cite{G,MUNIRO,GMMN}
\begin{equation}\label{2.15}
\left(\begin{array}{c}
u_{\ell}(R^+) \\  [2ex]  
u'_{\ell}(R^+)
      \end{array}
    \right)
=  
\left(
      \begin{array}{cc}
  \displaystyle      \frac{2+\beta}{2-\beta} & 0 \\[1ex]
\displaystyle \frac{4\alpha}{4-\beta^2} &
\displaystyle \frac{2-\beta}{2+\beta} \\
      \end{array}
    \right)
\left(
      \begin{array}{c}
u_{\ell}(R^-) \\  [2ex]  
u'_{\ell}(R^-)
      \end{array}
    \right), 
\quad \text{where}\quad     
    u_\ell(R^\pm) = \lim_{x\to R^\pm}u_\ell(x).
    \end{equation}

The function $u_{\ell}(r)$ is given by \eqref{2.8} and \eqref{2.13}. As already mentioned, there is a rigorous  discussion on the self adjointness of the resulting Hamiltonian   in appendix~\ref{A}.  
The matrix relation \eqref{2.15}, together with \eqref{2.8} and \eqref{2.13}, yields the following secular equation:
\begin{equation}\label{2.16} 
\frac{ \chi\, J_{\ell+\frac32}\left( \chi \right)}{J_{\ell+\frac12}\left(\chi \right)}=
 \frac{(2+\beta)^2}{(2-\beta)^2} 
\frac{\sigma\, K_{\ell+\frac32}\left(\sigma \right)}{K_{\ell+\frac12}\left(\sigma \right)} -\frac{8\beta(\ell+1)}{(2-\beta)^2}
+ \frac{w_{\ell j}}{(2-\beta)^2}.
\end{equation}
We will denote the left-hand side  by $\varphi_\ell(\chi)$ and the right-hand side\footnote{Observe that  $\phi_\ell(\sigma)$ also depends explicitly on $j$ and $\beta$.} by $\phi_\ell(\sigma)$, so \eqref{2.16} is written as $\varphi_\ell(\chi)=\phi_\ell(\sigma)$.
For simplicity, we have introduced the following auxiliary variables
\begin{equation}\label{3.1}
\chi= v_0 \sqrt{1-\varepsilon} \,,\qquad \sigma= v_0 \sqrt{\varepsilon}\,,
\end{equation}
and defined the  dimensionless parameters, $v_0$, $w_{\ell j}$ and the {\it relative energy} $\varepsilon$ as
\begin{equation}\label{2.17}
v_0= \sqrt{\frac{2\mu R^2 V_0}{\hbar^2}}>0\,,
\qquad 
w_{\ell j}= \frac{-8 \mu V_\text{so} \xi_{\ell j}R}{\hbar^2}\,,\qquad 
\varepsilon=|E|/V_0 \in (0,1).
\end{equation}

The secular equation \eqref{2.16} does not admit closed-form solutions for the energy of bound states and it will be analyzed in the forthcoming section. 

\section{General properties of the bound state structure}\label{section4}

In the previous section we have established the matching conditions that radial wave functions must fulfil so that the $\delta$ and the local $\delta'$ interactions are well defined. With this, in the present section  we consider the whole  Hamiltonian \eqref{2.2} in order to study the existence and properties of bound states.
 Moreover, in section \ref{sec.5.1} we  consider the cases for which the matching conditions are ill defined  and we give a simplified secular equation for large-parameter configurations in section~\ref{sec.5.2}.

\begin{figure}[h]
	\centering
\resizebox{0.3\textwidth}{!}{\includegraphics{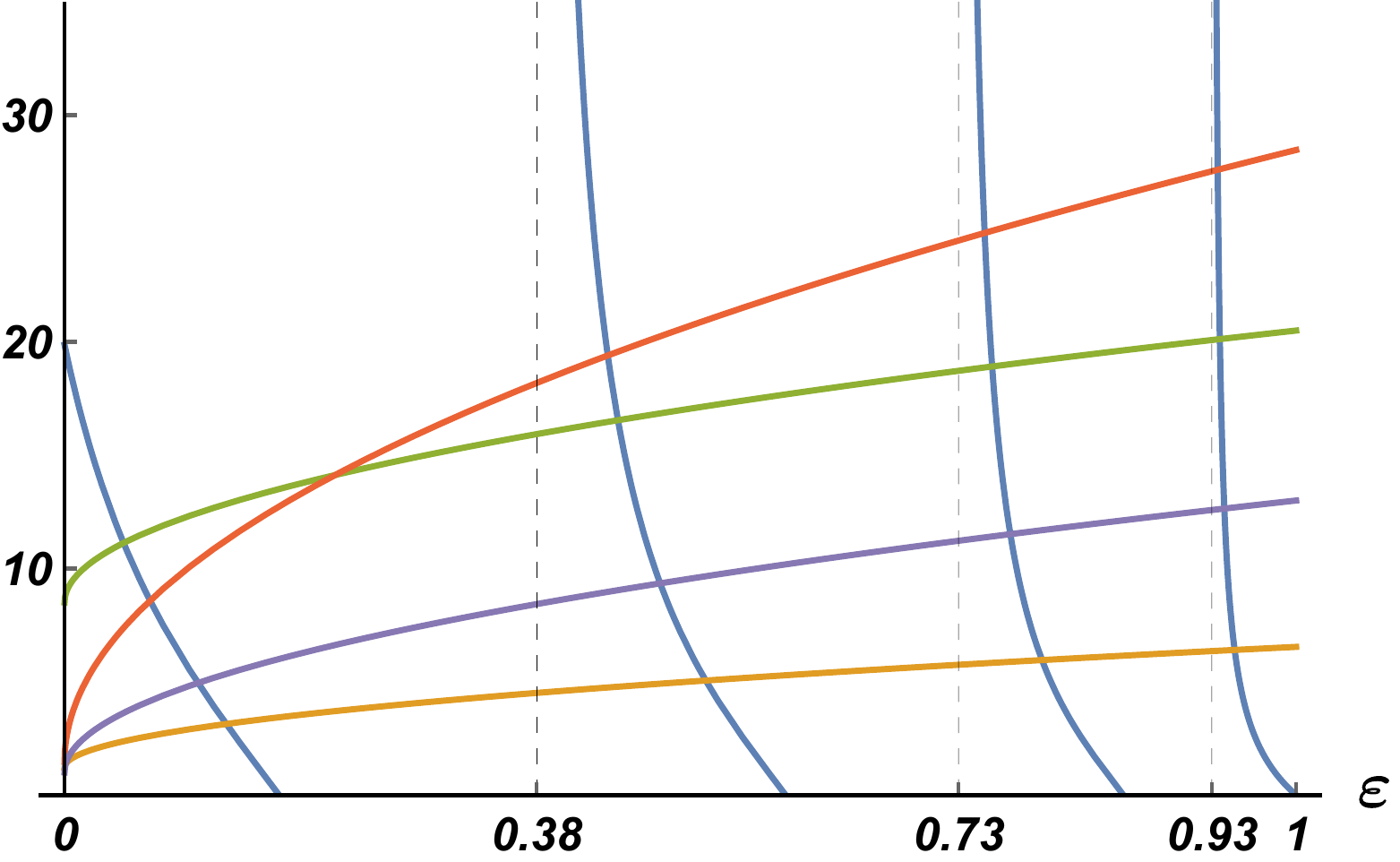}}
\qquad\qquad
\resizebox{0.3\textwidth}{!}{\includegraphics{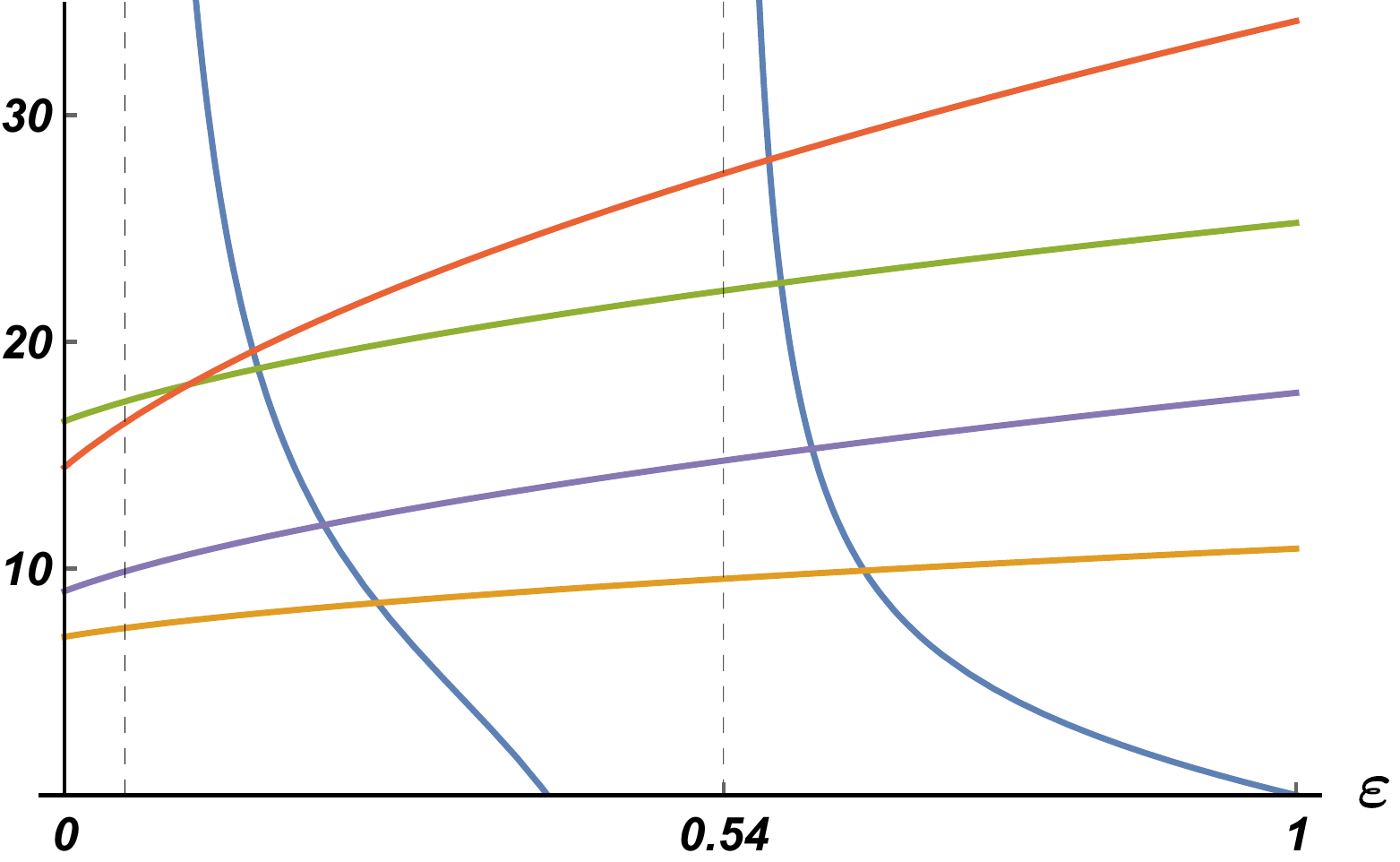}}
\caption{\small Bound states from the secular equation \eqref{2.16}: the left-hand $\varphi_\ell$ side in blue, the right-hand side $\phi_\ell$ in yellow for $\beta=-10$, green for $\beta=0$, red for $\beta=10$, and purple for $\beta=\pm\infty$. The figure on the left corresponds to $\ell=0$, the figure on the right to $\ell=4$. In both cases, the values of the relevant parameters are chosen to be $v_0=12$, $w_{\ell j}=30$.}
\label{fig:1}     
\end{figure}

Before proceeding with our presentation, let us denote by  $j_{\lambda, s}$ the $s$-th strictly positive zero of $J_\lambda(x)$, $\lambda>0$. As is well known, these zeros satisfy
\begin{equation}\label{3.7}
j_{\lambda,0}\equiv0<j_{\lambda,s}<j_{\lambda+1,s}<j_{\lambda,s+1}\,,\qquad s\in \mathbb{N}.
\end{equation}
We begin with a result concerning the existence and number of bound states,  whose proof is given in~appendix~\ref{B4}.
\begin{theorem}\label{prop:4}
If for any value $\ell \in\mathbb{N}_0$ such that $\ell\le \ell_\text{max}$ the following inequality holds
\begin{equation}\label{eq:Boundw_0}
w_{\ell j}> -\left((\beta -2)^2+2  \ell\left(\beta ^2+4\right)\right),
\end{equation}
there exists one, and only one, energy level with  relative energy 
\begin{equation}\label{3.8}
\varepsilon_s \in \left(1- \frac{j^2_{\ell+1/2,s}}{v_0^2}  , 1- \frac{j^2_{\ell +3/2,s-1}}{v_0^2} \right)\subset (0,1), \qquad s\in \mathbb{N}.
\end{equation}
In addition, for $w_{\ell j} \in \mathbb{R}$ the final number of bound states, $N_{\ell}=(2\ell+1)n_\ell$, is determined by 
 \begin{equation}\label{Prop4nl}
 n_{\ell}= M + m-m',
 \end{equation}
 where $M$ is 
 \begin{equation}\label{eq:M}
M=\min\{s\in \mathbb{N}_0 \,| \, j_{\ell+1/2,s+1}> v_0\},
\end{equation}
and, using the functions $\varphi_\ell(\chi)$ and $\phi_\ell(\sigma)$ defined after \eqref{2.16}, 
 \begin{eqnarray*}
m=\displaystyle\begin{cases}
 1  &\emph{if} \quad \varphi_\ell(v_0)> \phi_\ell(0^+),
 \\[1ex]
 0 &\emph{if} \quad \varphi_\ell(v_0)< \phi_\ell(0^+) \ \ \text{or}  \ \ v_0= j_{\ell+1/2,M} , \end{cases}
\qquad 
m'=\displaystyle\begin{cases}
 1  &\emph{if} \quad 0> \phi_\ell(v_0),
 \\[1ex]
 0 &\emph{if} \quad 0< \phi_\ell(v_0). \end{cases}
\end{eqnarray*}
\end{theorem}
Observe that the  cases $\varphi_\ell(v_0)=\phi_\ell(0^+)$ and $\phi_\ell(v_0)=0 $ are a priori excluded from the present study.
The same holds true for the possible bound states with energy below the potential well, which can arise if $\phi_\ell(v_0)<0$. We have focused on the states that are considered in mean-field nuclear models in order to facilitate the application.
In addition, it is interesting to point out that the structure of the  energy intervals \eqref{3.8} is unaffected by the $\delta'$ interaction as long as \eqref{eq:Boundw_0} holds. Moreover, the number of bound states is mainly determined by $M$.  For example, in Figure~\ref{fig:1} we observe that this number   remains the same for different values of $\beta$. The same conclusion holds for the isotope $^{209}$Pb, as will be shown in the last section.  This fact could eventually justify the interpretation of this $\delta'$ interaction as 
an extra mean-field interaction less relevant than the spin-orbit one.

In theorem~\ref{prop:4}, we have assumed the existence of an upper bound for the angular momentum, $\ell_\text{max}$. For  spherically symmetric potentials, satisfying $\int_0^\infty  |V(r)|\,r^t\, dr<\infty$ for $t=1,2$, the inequality 
\begin{equation}\label{eq:Bargmann}
n_\ell< \dfrac{1}{2 \,\ell +1}\int_0^\infty r |V_-(r)|\,dr,\quad  V_-(r)=\min(\{V(r),0\}),
\end{equation}
ensures  the existence of this upper bound \cite{BAR}.  For the $\delta$ potential alone, the existence of $\ell_\text{max}$ is guaranteed. As a matter of fact, it can be shown   that a particular linear combination of $\delta$ potentials  saturates the  previous inequality \cite{SCH}.
When we add the $\delta'$ term, the argument of Bargmann \cite{BAR} for the existence of $\ell_\text{max}$ does not apply any more, since it is then unclear how to interpret the previous integral. Fortunately, the following result,   for which the proof is given in~appendix~\ref{B5}, guarantees the existence of this bound in the present configuration. Furthermore, it also provides a simple expression for $\ell_\text{max}.$
\begin{theorem}\label{prop:5}
There are no  bound states with angular momentum  $\ell>\ell_{\max}$, where
\begin{equation*}
\ell_{\max}=\max\{\ell\in\mathbb{N}_0 \ | \ j_{\ell+1/2,1} <   v_0 \ \emph{or} \ \varphi_\ell(v_0)>\phi_\ell(0^+)\}.
\end{equation*}
If there exist $s_0 \in \mathbb{N}$ and  $\ell_0 \in \mathbb{N}_0$ such that $v_0= j_{\ell_0+1/2,s_0}$ the second  condition in the previous set can not be evaluated. Nonetheless, it is not necessary since the existence of at least one bound  state for $\ell_0$  is guaranteed.
\end{theorem}
Concerning the ordering of bound states, we can prove an important result, that will be useful in the sequel. For spherically symmetric potentials, we know that if $E_{n\ell}$ denotes the energy of a bound state defined by the quantum numbers $n$ and $\ell$ the following inequalities hold,
\begin{equation}\label{EEE}
E_{n\ell}<E_{(n+1)\ell}<E_{(n+1)(\ell+1)},\quad n,\ell \in \mathbb{N}_0.
\end{equation}
This statement can be derived for continuous potentials using Sturm's theorem analysing the spectral properties of the Hamiltonian \cite{GA}. Now, we extend this result for the spherically symmetric $\delta$-$\delta'$ interaction we are dealing with \eqref{2.2}, where we have to take into account  an additional quantum number $j$. The proof of the following result is given in~appendix~\ref{B6}.
\begin{theorem}\label{prop:6} If there exist  bound states with relative energies $\varepsilon_{n\ell_j}, \varepsilon_{(n+1)\ell_j}, \varepsilon_{n(\ell+1)_j}$ for $n,\ell \in \mathbb{N}_0$ the following inequalities hold:
\begin{eqnarray}\label{eq:pro3}
 (a)\ -\varepsilon_{n\ell_j}<-\varepsilon_{(n+1)\ell_j} , \qquad  
 (b)\ -\varepsilon_{n\ell_{j}}<-\varepsilon_{n(\ell+1)_{j}},  \qquad
 (c)\ -\varepsilon_{n\ell_{\ell + 1/2}}<-\varepsilon_{n\ell_{\ell - 1/2}}. 
\end{eqnarray}
The second inequality only applies for $j=\ell+1/2$ and the third inequality   for $\ell>0$.
\end{theorem} 
These theorem is in agreement with the results of the nuclear shell model, reinforcing the application as a limiting case of the Woods-Saxon potential in section \ref{sec.isotopes}.  Inequalities (a) and (b) are well known in nuclear physics  when dealing with spherically symmetric potentials \cite{{Mayer,Bertsch}}, as it was already mentioned in \eqref{EEE}. With respect to inequality (c), it is worth mentioning that the microscopic quantum description of nucleons inside the nuclei, requires a careful treatment of
the orbital angular momentum with the intrinsic nucleon spin. This was connected with the long-standing problem of the inability to theoretically explain the magic
numbers in atomic nuclei. Only when this interaction was included in the mean-field shell model, all
experimental magic number were explained. In the course of this breakthrough it was found that,
contrary to atomic electrons, the nucleon which is aligned with the orbital angular momentum is
more strongly attracted. This is consistent with the previous theorem.

Lastly, a brief comment on the ground state. For a single particle Schr\"odinger equation it can be shown, using the variational principle,  that the ground state must be a  spherically symmetric zero angular momentum state \cite{Petit}. For the  spherical potential well, this statement can be directly proved with the secular equation \eqref{2.16} using the monotonicity properties of $\varphi_\ell(\chi(\varepsilon))$ and $\phi_\ell(\sigma(\varepsilon))$; in this case $\alpha=0$ and $\beta=0$ so the right-hand of equation \eqref{2.16}  is strictly positive. 

However, with the results shown, this could seem to be  no longer true when we add the $\delta$-$\delta'$ interaction. In fact, it would be enough to include the $\delta$ potential. For example,   there exists configurations with bound states for $\ell=1$ and not for $\ell=0$. 
Although  a  two particle system with strong spin-orbit coupling can end up with a non-zero angular momentum ground state \cite{GA2}, in  the above-mentioned configurations   the $\ell=0$ bound state exits. As we have already mentioned after theorem~\ref{prop:4}, an  attractive $\delta$ coupling such that $\phi_\ell(v_0)$ involves a bound state with energy below $-V_0$,  whereas we are focusing on states lying within  $(-V_0, 0)$,  see section \ref{sec:inside}. It is worth mentioning that the $\delta$-$\delta'$ interaction without the spherical  well also presents bound states with angular momentum $\ell\in\{0,1,\dots,\ell_\text{max}\}$. This statement has been proved in theorem~2 of \cite{MUNIRO}.

\subsection{Special cases $\beta=\pm 2$}\label{sec.5.1}

Let us go back to the matching conditions \eqref{2.15}. They do not apply for the exceptional values $\beta=\pm 2$. Nevertheless, there exist respective self adjoint extensions of the radial Hamiltonian for these cases \cite{KU}. They are  characterized by the following BC at $r=R$:
\begin{eqnarray*}
u_\ell(R^+)-\frac 4\alpha\,u'_\ell(R^+)=0, && \quad u_\ell(R^-)=0,\quad {\rm if} \quad \beta=2\,, \nonumber\\[1ex]  
u_\ell(R^-)+\frac 4\alpha\,u'_\ell(R^-)=0, && \quad u_\ell(R^+)=0,\quad {\rm if} \quad \beta=-2\,.\label{4.7}
\end{eqnarray*}
These situations have already been studied, for instance in \cite{GMMN}, where it is shown that in both cases the contact interaction becomes an opaque barrier, which means that the transmission coefficient is equal to zero. This suggests that there are only bound states, in an infinite number, plus scattering states and no resonances whatsoever. We may give an estimation of the values and number of bound states when $\beta\to\pm 2$. 

First of all, taking the limit $\beta\to 2$ in \eqref{2.16} when $\text{sgn}(\phi_\ell(0^+)\phi_\ell(v_0))=1$ we obtain $|\phi_\ell(\sigma)|\to\infty$. Therefore, the acceptable values of $\chi$ in the same equation are, essentially, the zeros of $J_{\ell+1/2}(\chi)$. Hence, 
from $\chi=v_0\sqrt{1-\varepsilon}$, we conclude that the $i$-th bound state with relative energy $\varepsilon_i$ is given by
\begin{equation}\label{4.9}
\lim_{\beta\to 2}\varepsilon_i = 1-\left(\frac{j_{\ell+1/2,i}}{v_ 0}\right)^2.
\end{equation}
Note that the first energy value, $\varepsilon_1$, is not reached if $\text{sgn}(\phi_\ell(0^+))=-1$. 
Secondly,  in the limit  $\beta\to -2$ equation \eqref{2.16} reduces to
\begin{equation}\label{zzz} 
\frac{ \chi\, J_{\ell+\frac32}\left( \chi \right)}{J_{\ell+\frac12}\left(\chi \right)}
=  \left(\ell+1 + \frac{w_{\ell j}}{16} \right).
\end{equation}
This transcendental equation is far simpler than \eqref{2.16} since the right-hand side is  independent of $v_0$ and the relative energy. 

\subsection{Large-parameter configurations}\label{sec.5.2}

The main aim of this section is to show that for certain values of the parameters $v_0, w_{\ell j}$ and $\beta$ remarkable simplifications in the bound state structure occur. To begin with, let us  consider $v_0\gg 0$.
Hence, using the limiting forms of the Bessel functions for large values of their arguments \cite{OLVER}, the secular equation (\ref{2.16}) can be approximated by
\begin{equation}\label{SecularLimiting}
\chi \cot \left(\frac{\pi  \ell}{2}-\chi\right)= \frac{(\beta +2)^2 \sigma-8 \beta  (\ell+1) +w_{\ell j}}{(\beta -2)^2}.
\end{equation}
It is important to note that the the previous equation  only differs from the zero angular momentum secular equation   by the term $8\beta(\ell+1)= 8\beta$. Consequently, for the $\delta$ potential alone and $\ell=0$,  equation \eqref{2.16} 
 takes the following simple form:
 \begin{equation}\label{3.3}
-\chi  \cot \chi = \dfrac{w_{\ell j}}{4}+\sigma\,.
\end{equation}
 In this regard, we should mention that this approach is valid for low angular momentum values only. For instance, after \eqref{SecularLimiting} we can not conclude the existence of the maximal angular momentum $\ell_\text{max}$ defined in theorem~\ref{prop:5}. 

If, in addition,   we consider $|w_{\ell j}|\gg |\beta|$  so that the right-hand side of \eqref{SecularLimiting} is nearly independent of $\varepsilon$, the energy of the bound states can be obtained, in an approximate form, from the zeros of $\sin\left(\frac{\pi  \ell}{2}-v_0\sqrt{1-\varepsilon } \right)$,\textit{ i.e.},
\begin{equation}\label{NegativeEnergyValues}
\varepsilon_n\simeq 1-\left( \frac{\pi (\ell-2 n)}{2 v_0}\right)^2\in(0,1),\qquad n\in \mathbb{Z}.
\end{equation}
Then,  we may estimate the number  of bound states for a given value $\ell$ of the angular momentum as
\begin{equation} \label{eq.nl}
N_\ell= n_\ell (2\ell +1)\quad \text{with}\quad n_\ell\simeq\left\lfloor\dfrac{\pi  \ell+2\, v_0}{2 \pi } \right\rfloor\simeq\left\lfloor\dfrac{v_0}{\pi}\right\rfloor,
\end{equation}
where the number of negative energy values $n_\ell$ has been obtained from (\ref{NegativeEnergyValues}), under the condition $\varepsilon_n>0$.  

Finally, irrespective of the previous considerations, we analyse a system characterized by a very strong $\delta'$ interaction, that is to say, we take the limit $|\beta|\to\infty$ in the secular equation \eqref{2.16}. As can be easily checked, this situation is equivalent to the non-existence of the $\delta$-$\delta'$ interaction, \textit{i.e.},  
$\alpha=\beta=0$. For this particular example, the matching conditions \eqref{2.15} impose the continuity of the radial function and its first derivative. The resulting secular equation   matches with the  one found for the finite three-dimensional spherical potential well, usually derived imposing the continuity of the logarithm derivative of the radial function at $R$ \cite{GA}. 

\section{Resonances}\label{section5}

Solvable or quasi-solvable models usually have, in addition to bound states, resonances (unstable or quasi-stable quantum states) and possibly anti-bound states. The study of resonance models is necessary because most of the known quantum states are unstable. 
For example, single-particle resonances appear in the dripline of light nuclei, such as ${}^5$He, ${}^8$B, and  ${}^{10}$Li. 
Resonance models give a qualitative account for resonance behaviour and, therefore, may give a good insight into the quantum properties of  unstable states. 
In this paper, we are assuming that resonances appear in resonance scattering, which is produced by a Hamiltonian pair $\{H_0,H=H_0+V\}$. Thus, a resonance arises when the incoming particle stays  in the region where the potential acts a  much longer time than the one it would have stayed if the potential had not existed.

There are several definitions of resonances based on either physical or mathematical notions, which are not always equivalent. Because of the kind of model presented here, we are using the concept of resonance as given in mathematical terms. There are essentially two approaches, either we define resonances as poles of analytic continuations of a \textit{reduced} resolvent of the total Hamiltonian \cite{RSIV}, or  as poles of an analytic continuation of the $S$-matrix in the momentum representation (or equivalently in the energy representation). Here, we shall adopt the second point of view.

Under some general conditions based on causality principles \cite{NUS}, the $S$ matrix in momentum representation, denoted by $S(k)$, admits an analytic continuation to a meromorphic function of the complex variable $k$ on the whole complex plane. It is meromorphic because $S(k)$ has poles, which may be classified in three types: 
\begin{itemize}
	\item 
	Simple poles on the positive half of the imaginary axis that correspond to bound states.
	\item 
	Simple poles in the negative half of the imaginary axis, which represent the presence of the antibound (virtual) states.
	\item 
	Pair of poles on the lower half plane, symmetrically located with respect to the imaginary axis, each of these pairs representing one resonance \cite{NUS,BOH}. 
\end{itemize}
Although the order of resonance poles may be in principle arbitrary (both poles of each pair must have the same multiplicity), in general they are simple. This  result emerges from our particular model.

If we go from the momentum to the energy representation, $E={\hbar^2k^2}/(2\mu)$, poles for each resonance pair become two conjugate complex numbers of the form $z_R=E_R-i\Gamma/2$ and $z_R^*=E_R+i\Gamma/2$, with $\Gamma>0$.  Here, $E_R$ represents the resonance energy, usually $E_R>0$, and $\Gamma$ the inverse of the half life. After this, one may understand that in the momentum representation, the closer a resonance pole is to the real axis, the higher is its mean life.

Without further ado, let us study the resonances in the present case. In section \ref{sec:outside}, we have written the wave equation outside the nucleus as a linear combination of modified Bessel functions of first and second kind. The requirement of square integrability, needed to characterize bound states, forced us to drop the contribution of the Bessel function of first kind and just keep the Bessel function of second kind. 
Resonance state functions, also called Gamow functions, are not square integrable so we can simply solve Schr\"odinger equation at $r>R$ for $E>0$. This leads to a solution analogous to \eqref{2.6} in terms of Bessel and Neumann functions.
Nevertheless, for reasons that will be evident  below, it is convenient to write this solution in terms of the H\"ankel functions as
\begin{eqnarray}\label{5.1}
 u_\ell(r)=\sqrt{\kappa  r} \left(C_\ell H_{\ell+\frac{1}{2}}^{(1)}(\kappa  r) + D_\ell H_{\ell+\frac{1}{2}}^{(2)}(\kappa  r)\right),  \qquad \kappa = \frac{\sqrt{2{\mu}E}}{\hbar}\,,\quad E>0\,,
\end{eqnarray} 
where $C_\ell$ and $D_\ell$ are independent of $r$, although they depend on $\kappa$. This expression is valid for $r>R$.  The superscripts distinguish between the H\"ankel functions of first and second kind.  These  functions present the following asymptotic behaviour \cite{OLVER} for large values of $r$:
\begin{eqnarray*} 
H_{\ell+\frac{1}{2}}^{(1)}(\kappa r) \sim \sqrt{\frac{2}{\pi\,\kappa r}}\; e^{i(\kappa r-(\ell+1)\pi/2)}\,, \quad  -\pi<\arg z<2\pi\,,
\qquad 
H_{\ell+\frac{1}{2}}^{(2)}(\kappa r) \sim \sqrt{\frac{2}{\pi\,\kappa r}}\; e^{-i(\kappa r-(\ell+1)\pi/2)}\,.
\end{eqnarray*}
Consequently, $H_{\ell+ {1}/{2}}^{(1)}(\kappa  r)$ can be interpreted as an outgoing wave function, while $H_{\ell+ {1}/{2}}^{(2)}(\kappa  r)$ as an incoming wave function. Resonances are given by the so called {\it purely outgoing boundary condition}, which states that only the outgoing wave function survives. This is   satisfied if, and only if, $D_\ell=0$ in \eqref{5.1}. At first look, this may resemble to the requirement $C_\ell=0$ for \eqref{2.11}, although the situation here has a completely different origin. Furthermore, the transcendental equation $D_\ell(\kappa)=0$ gives us the poles of the $S$-matrix, which are the resonant poles.

\begin{figure}[h]
	\centering
	\resizebox{0.3\textwidth}{!}{
		\includegraphics{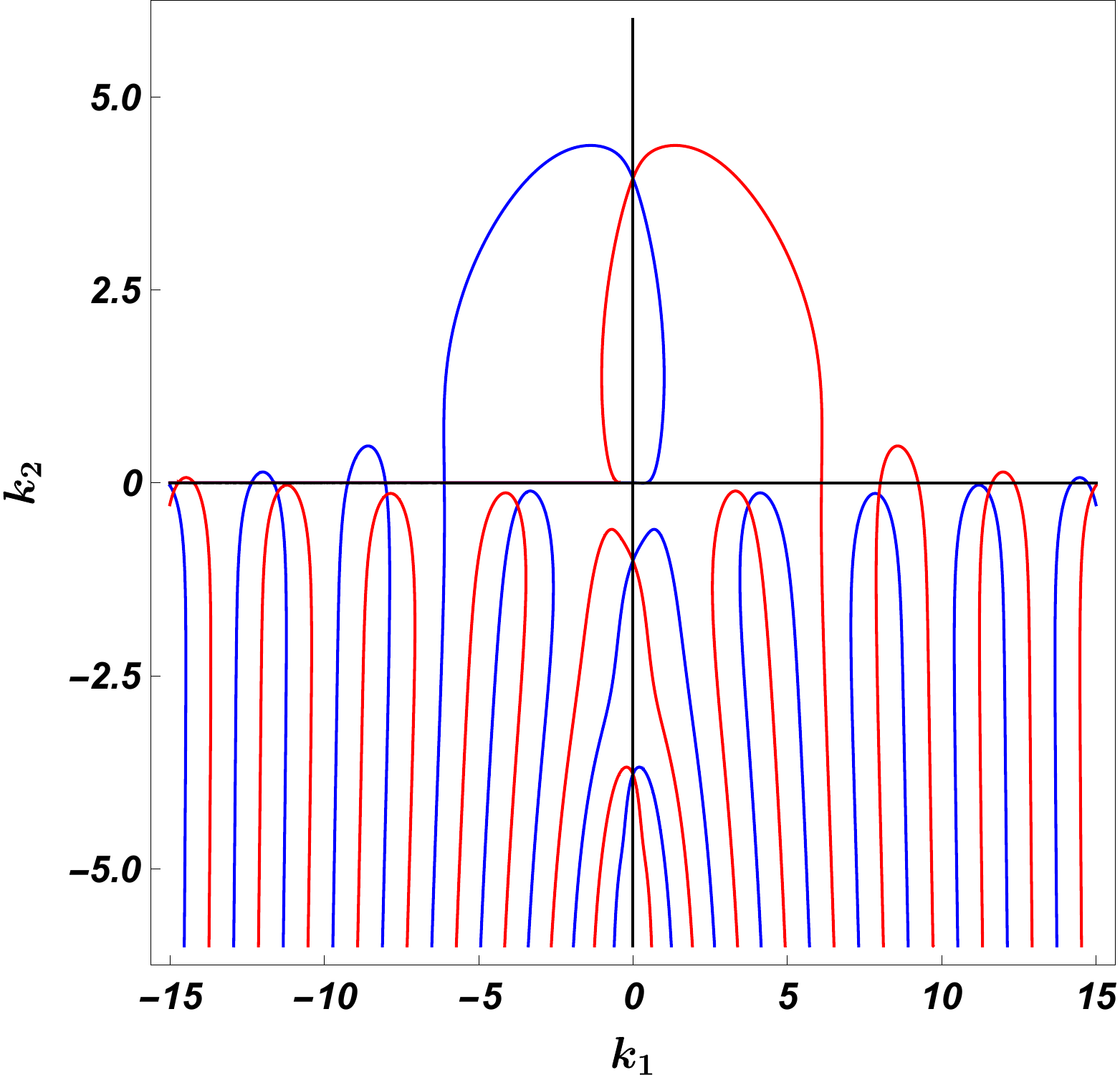}
	}
	\caption{\small In blue ${\rm Re}\, F (k_1,k_2) =0$ and in red ${\rm Im}\, F (k_1,k_2) =0$, from \eqref{5.4} for $\ell=0$. Bound states and resonances correspond to intersection of red and blue curves. The relevant parameters are chosen to be $v_0=5$, $w_{\ell j}=10$ and $\beta=1$.}
	\label{fig_3}       
\end{figure}

In order to obtain these poles in the momentum representation and, after that, proceed with the construction of the resonance Gamow wave functions, we also use the matching condition  between the outgoing function and the wave function inside the potential well,  previously calculated in section \ref{sec:inside}. This gives the following transcendental equation in $k$
\begin{eqnarray}
&&H^{(1)}_{\ell+\frac12} (R\kappa) \bigl[ 8(\alpha R-\beta) J_{\ell+\frac12}(R\gamma)-(\beta-2)^2 R\gamma J_{\ell+\frac32}(R\gamma)   +  (\beta-2)^2 R\gamma J_{\ell-\frac12}(R\gamma)\bigr] \nonumber   \\ 
&& \qquad\qquad\quad  + (\beta+2)^2 \kappa R J_{\ell+\frac12}(R\gamma) H^{(1)}_{\ell+\frac32} (R\kappa)    - (\beta+2)^2 \kappa R J_{\ell+\frac12}(R\gamma) H^{(1)}_{\ell-\frac12} (R\kappa) =0\,.  \label{5.4} 
\end{eqnarray}
The solutions of \eqref{5.4} should be classified in three categories, as previously explained.
If we set $\ell=0$, the situation simplifies enormously. In fact, \eqref{5.4} becomes
\begin{equation}\label{5.5}
\frac{\tan(\gamma R)}{\gamma R} =  -\frac{i(\beta-2)^2}{(\beta+2)^2\, \kappa R+4i\alpha R}\,.
\end{equation}
When we choose $\beta=0$ (absence of the term in $\delta'$), $V_0=0$ ($\gamma=\kappa$) and $\ell=0$, we recover well known results of one-dimensional systems \cite{DO,GNN}.
Since the resonant poles are complex solutions in the momentum representation, let us use the following notation:
\begin{equation}\label{5.6}
k_1+ik_2 =: R\kappa\,,\quad \quad R\gamma= \sqrt{v_0^2+(k_1+ik_2)^2}\,,
\end{equation}
so that \eqref{5.5} may be written as
\begin{eqnarray}\label{5.8}
F(k_1,k_2) &:=& \frac{\tan \sqrt{v_0^2+(k_1+ik_2)^2}}{\sqrt{v_0^2+(k_1+ik_2)^2}} + \frac{i(\beta-2)^2}{(\beta+2)^2 (k_1+ik_2)+iw_{\ell j}} =0\,. 
\end{eqnarray}
Denoting the real and imaginary parts of the complex number $z$ by ${\rm Re}\,z$ and ${\rm Im}\,z$, respectively, a simple analysis on \eqref{5.8} shows that
\begin{eqnarray}
{\rm Re}\, F (-k_1,k_2) = {\rm Re}\, F(k_1,k_2)\, \quad {\rm and}  \quad
-{\rm Im}\, F(-k_1,k_2) ={\rm Im}\,F(k_1,k_2)\,.  \label{5.9}
\end{eqnarray}
We observe that \eqref{5.9} implies that the curves in the plane $(k_1,k_2)$ given by ${\rm Re}\, F(k_1,k_2)=0$ and ${\rm Im}\,F(k_1,k_2)=0$ are symmetric with respect to the imaginary axis $k_1=0$. 
The behaviour of the solutions of equation \eqref{5.8} is shown in Figure~\ref{fig_3}. Except for the intersections of these two curves in the negative imaginary semi-axis,  antibound states, the ones in the lower half plane give the resonance poles. Two intersections symmetrically placed with respect to the imaginary axis $k_1=0$ give the same resonance. 

\begin{figure}[hb]
	\centering
	\resizebox{0.3\textwidth}{!}{
		\includegraphics{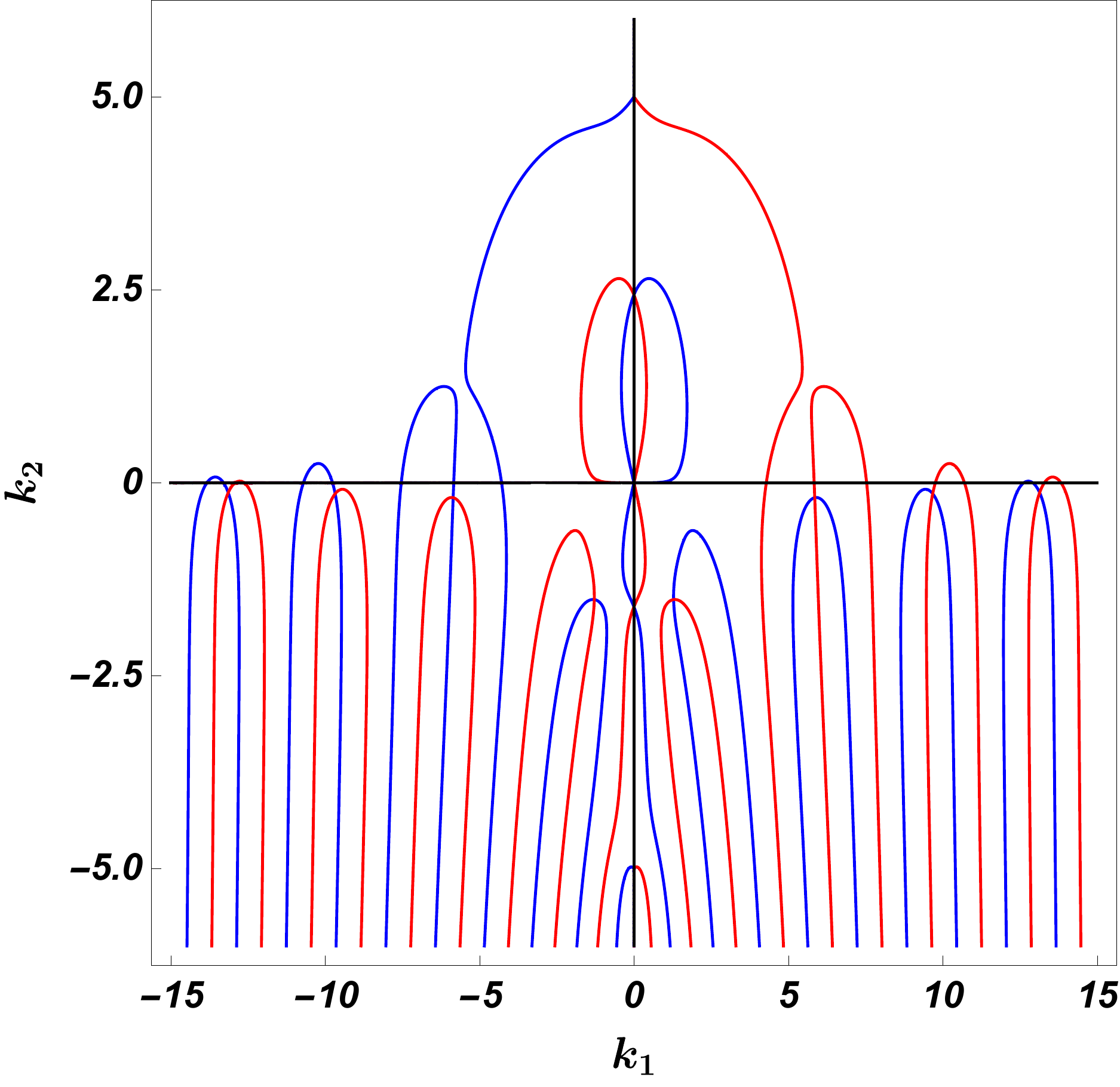}}
	\qquad 
	\resizebox{0.3\textwidth}{!}{
		\includegraphics{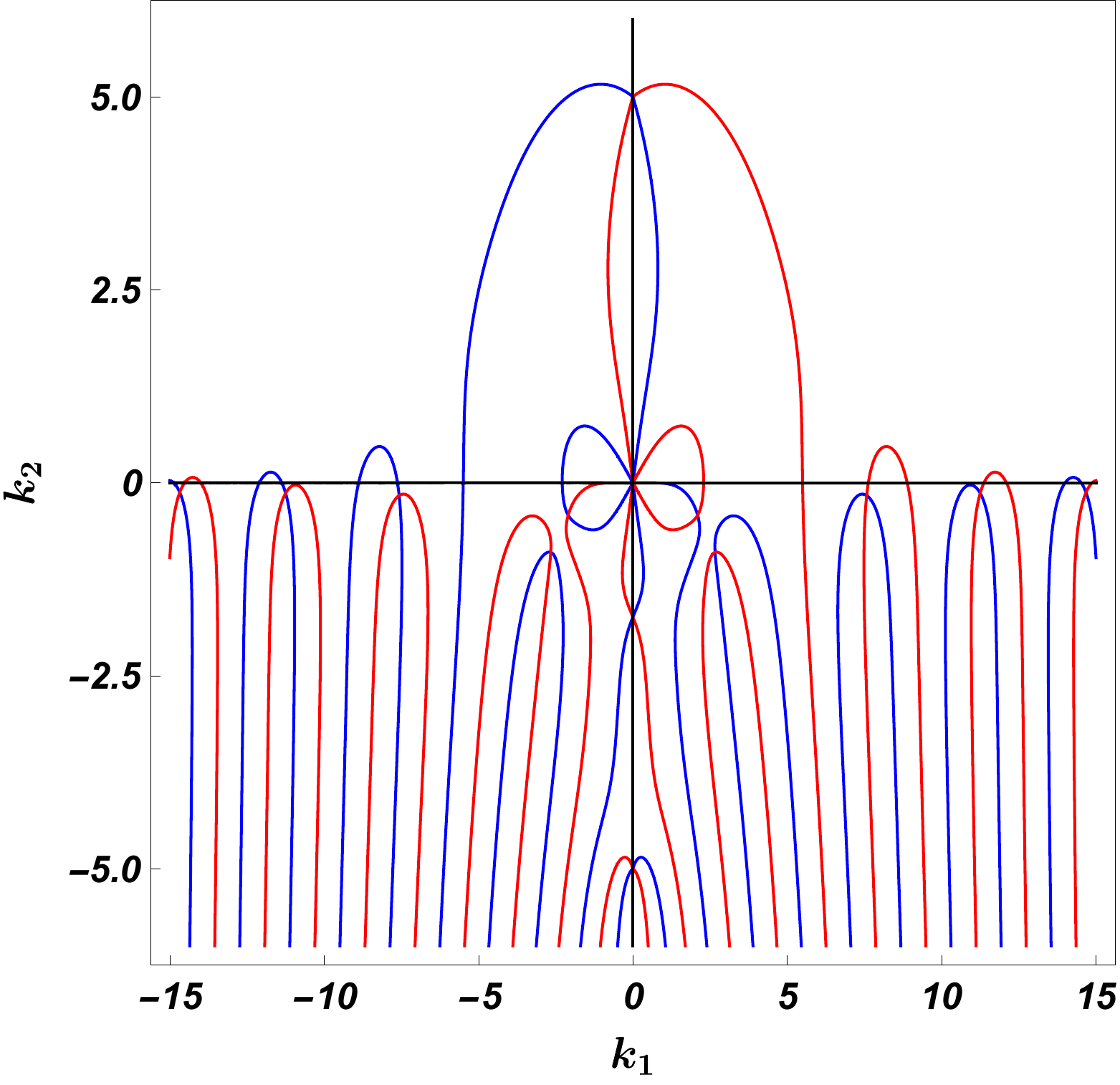}} 
	\\
	\resizebox{0.3\textwidth}{!}{
		\includegraphics{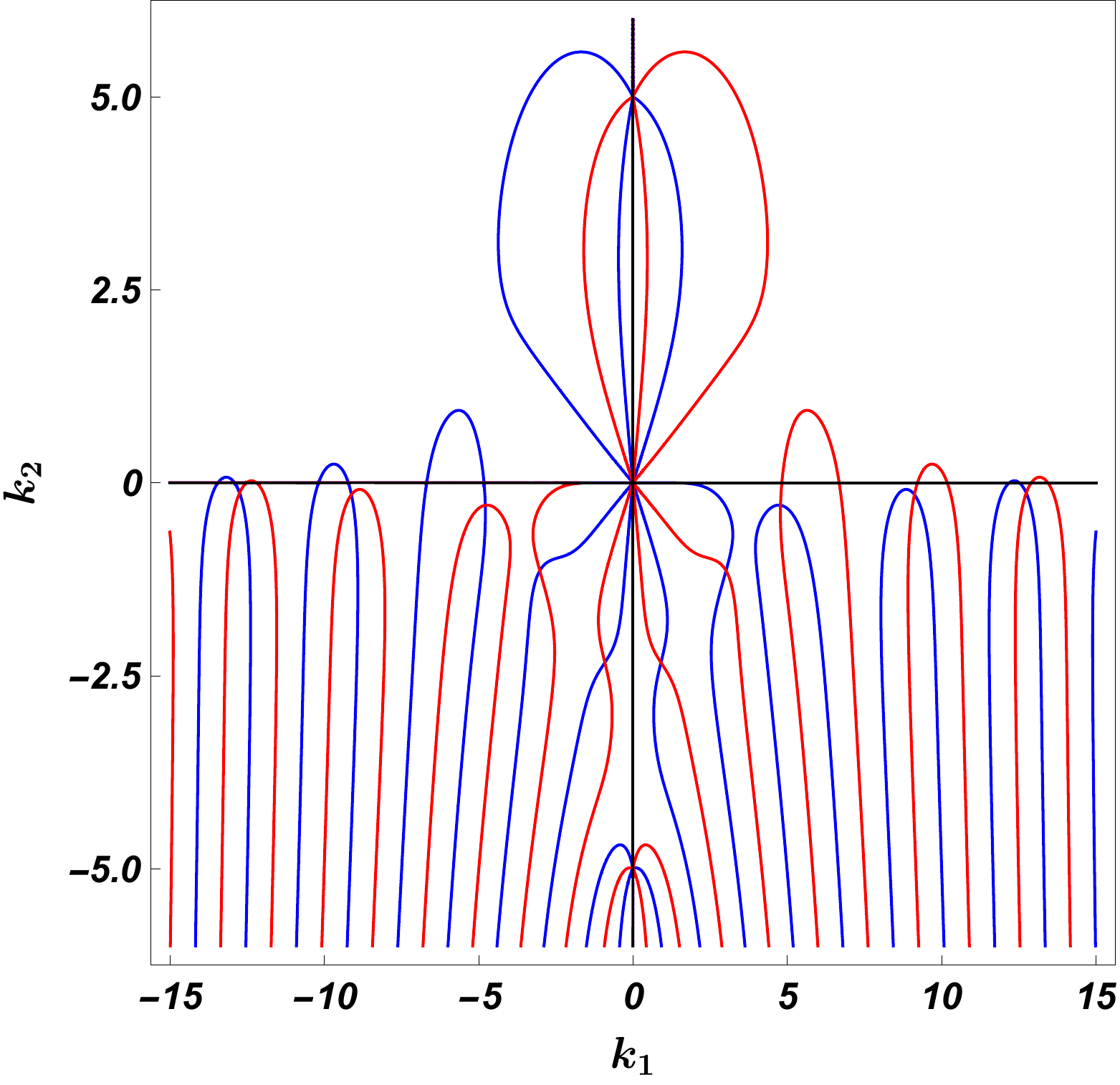}}
	\qquad
	\resizebox{0.3\textwidth}{!}{
		\includegraphics{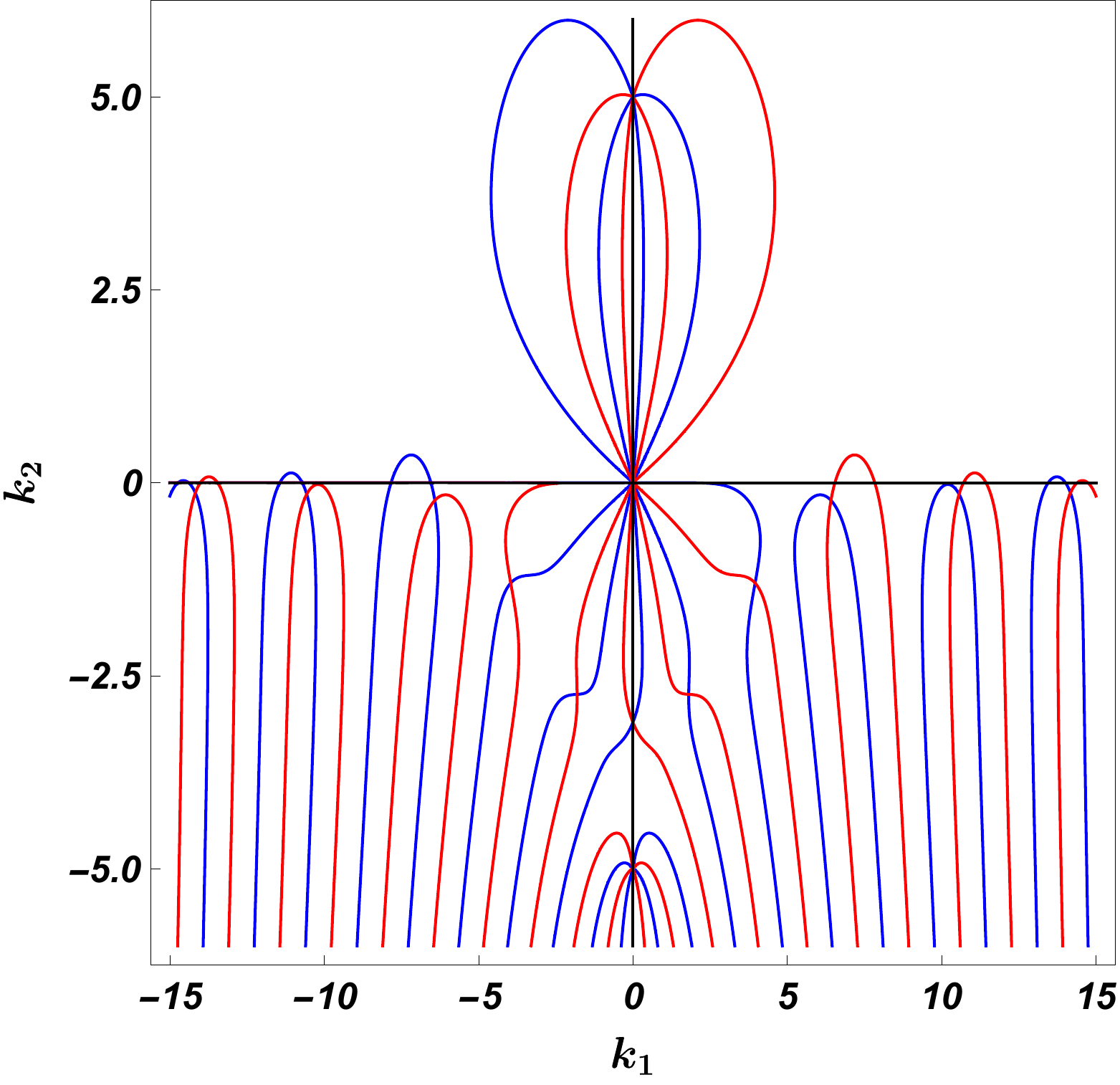}}  
	\caption{\small
	Annihilation of \eqref{5.4}: in blue the real part, in red the imaginary part. Bound states and resonances correspond to the intersection of red and blue curves. From top to bottom and left to right the curves for $\ell=1, 2, 3, 4$. The relevant parameters are chosen to be $v_0=5$, $w_{\ell j}=10$ and $\beta=1$.}
	\label{fig_4}       
\end{figure}
We also observe the existence of a bound state in the positive imaginary axis $k_2>0$. These results are consistent with those obtained  for bound states earlier in this paper;  when we set $k_1=0$ and $k_2>0$ in \eqref{5.8} we recover the secular equation  for $\ell=0$, see \eqref{SecularLimiting} and the comment underneath. 
It is noteworthy that there is an infinite number of resonances which lie on the lower half plane without the real axis $k_2=0$. In fact, for $k_2=0$, the imaginary part of \eqref{5.8} is given by
\begin{equation}\label{5.10}
\frac{(\beta^2-4)^2 k_1}{(\beta+2)^4 k_1^2+w_{\ell j}^2}=0\,,
\end{equation}
which implies  $k_1=0$, so that all intersections should coincide on the origin, which is obviously not the case. This is important, since as a consequence of reasonable {\it causality conditions}  \cite{NUS} resonance poles should lie on the lower half plane in the momentum representation.

Equation \eqref{5.4} for $\ell\ne 0$ does not admit the kind of simplification yielding to \eqref{5.5}. Nevertheless, it is still possible to give an estimation of the location of the first few resonances in the $k$ plane as well as some antibound states.  Our results are depicted in Figure~\ref{fig_4}, where the cases $\ell=1,2,3,4$ are considered. 
Resonance poles are located at the intersections of red and blue curves right below the real axis. Antibound  states poles are located at the intersections of blue and red curves on the negative imaginary axis.  The structure of the solutions is similar to the case $\ell=0$.

\section{Neutron energy levels of ${}^{\textbf{133}}$Sn and ${}^{\textbf{209}}$Pb} \label{sec.isotopes}

The purpose of this final section is to briefly discuss the general results previously obtained in the context of a realistic physical situation. 
To begin with, we are going to use the program know as the {\it Gamow code} for some of our estimations. It is a numerical program which gives the energy of bound states for the Woods-Saxon potential \eqref{eq.h2} and it is quite useful for various reasons. First of all, it serves to estimate how good  the approximation $a\to 0^+$ is. In addition, the Gamow code permits a comparison with the experimental results. Note that this code supplies more values than the current experimental data. 
We should also remark that our goal is to show that our results are qualitatively reasonable for low-lying bound states and that we do not intend to get a numerical fit with good precision, which is beyond the purpose of the model. 

In Table~\ref{table.eexp}, we compare the Gamow code   \cite{1982Vertse} and experimental energies, taken from the Database of the National Nuclear Data Center  Brookhaven National Laboratory \cite{nndc} and \cite{Bromley}, for the isotope ${}^{209}$Pb. With this comparison we ascertain that the program we are using to test our model fits with the available experimental data. The relevant parameters describing the lowest experimental energy states are 
$V_0 = 44.4$~MeV, 
$V_\text{so}=16.5$~MeV~fm, 
$r_0=1.27$~fm ($R=7.525$~fm), 
 $a=0.7$~fm, 
and
${2 \mu}/{\hbar^2}=0.0480$~MeV$^{-1}$~fm$^{-2}$ \cite{nndc}. 	
For the present configuration 
$v_0=10.98,\ w_{\ell j}=-23.83 \,\xi_{\ell j}$.
\begin{table}[h!]
	\centering
	\caption{Comparison of the numerical (Gamow code)  energy levels (MeV)  in $^{209}$Pb with the experimental ones, using the physical parameters mentioned in the text.}
	\label{table.eexp}      
	\begin{tabular}{lll}
		\hline\noalign{\smallskip}
		State ($n\ell_j$) & Gamow  & $E_\text{exp}$  \\
		\noalign{\smallskip}\hline\noalign{\smallskip}
		$1g_{9/2}$  &   -3.93                 & -3.94   \\
		$2p_{1/2}$  &    -7.41                & -6.73     \\
		$2p_{3/2}$  &    -8.35                 & -7.62   \\
		$0i_{11/2}$  &  -2.80                   & -3.16   \\
		$2d_{5/2}$  &  -2.07                   &  -2.37    \\
		$0j_{15/2}$  &  -1.88                   &   -2.51 \\
		\noalign{\smallskip}\hline
	\end{tabular}
\end{table}

Focusing on the results that emerge from the singular Hamiltonian \eqref{2.2}, we begin  comparing the energy levels for some states of the nucleus $^{133}$Sn achieved using the square well plus the $\delta$ potential alone, $\beta=0$.   The energy values of the $\delta$-$\delta'$ model ($\delta$-$\delta'$ M) are obtained through \eqref{2.16}, where the numerical values of the physical parameters  are  $V_0=39.5$~MeV, $V_\text{so}=15.5$~MeV~fm, $r_0=1.27$~fm, ${2 \mu}/{\hbar^2}=0.0479~\text{MeV}^{-1}~\text{fm}^{-2}$, $R=6.47$~fm  and $V_\text{q}=0$ \cite{nndc}, which set $v_0=8.89$, $w_{\ell j}=-19.20 \,\xi_{\ell j}$, see \eqref{2.17}. Some results are shown in Table~\ref{table.sn}, where we can see that the inequalities of \eqref{eq:pro3} are always satisfied. For these low-lying bound states we obtain a quantitatively fair approximation. For this simulation we have taken $a=0.05$ fm.
\begin{table}[h!]
\centering
\caption{\small Neutron energy levels (MeV) for $\ell=0$ and $\ell=1$ in the nucleus $^{133}$Sn, using the physical parameters mentioned in the text.}
\label{table.sn}       
\begin{tabular}{lll}
\hline\noalign{\smallskip}
State ($n\ell_j$) & Gamow & $\delta$-$\delta'$~M  \\
\noalign{\smallskip}\hline\noalign{\smallskip}
$0s_{1/2}$ & -35.52  & -35.53  \\ 
$1s_{1/2}$ & -23.81  & -23.83 \\
$2s_{1/2}$ & -5.56  &  -5.59 \\[0.25ex]
\hdashline\\[-2.25ex]
$0p_{1/2}$  & -31.37  &   -30.79\\
$1p_{1/2}$  &  -15.95  &   -14.20\\
$0p_{3/2}$  &  -31.42  &   -31.95\\
$1p_{3/2}$  & -16.08  &   -17.46\\
\noalign{\smallskip}\hline
\end{tabular}
\end{table}

\begin{figure}[h]
\centering\resizebox{0.3\textwidth}{!}{
 \includegraphics{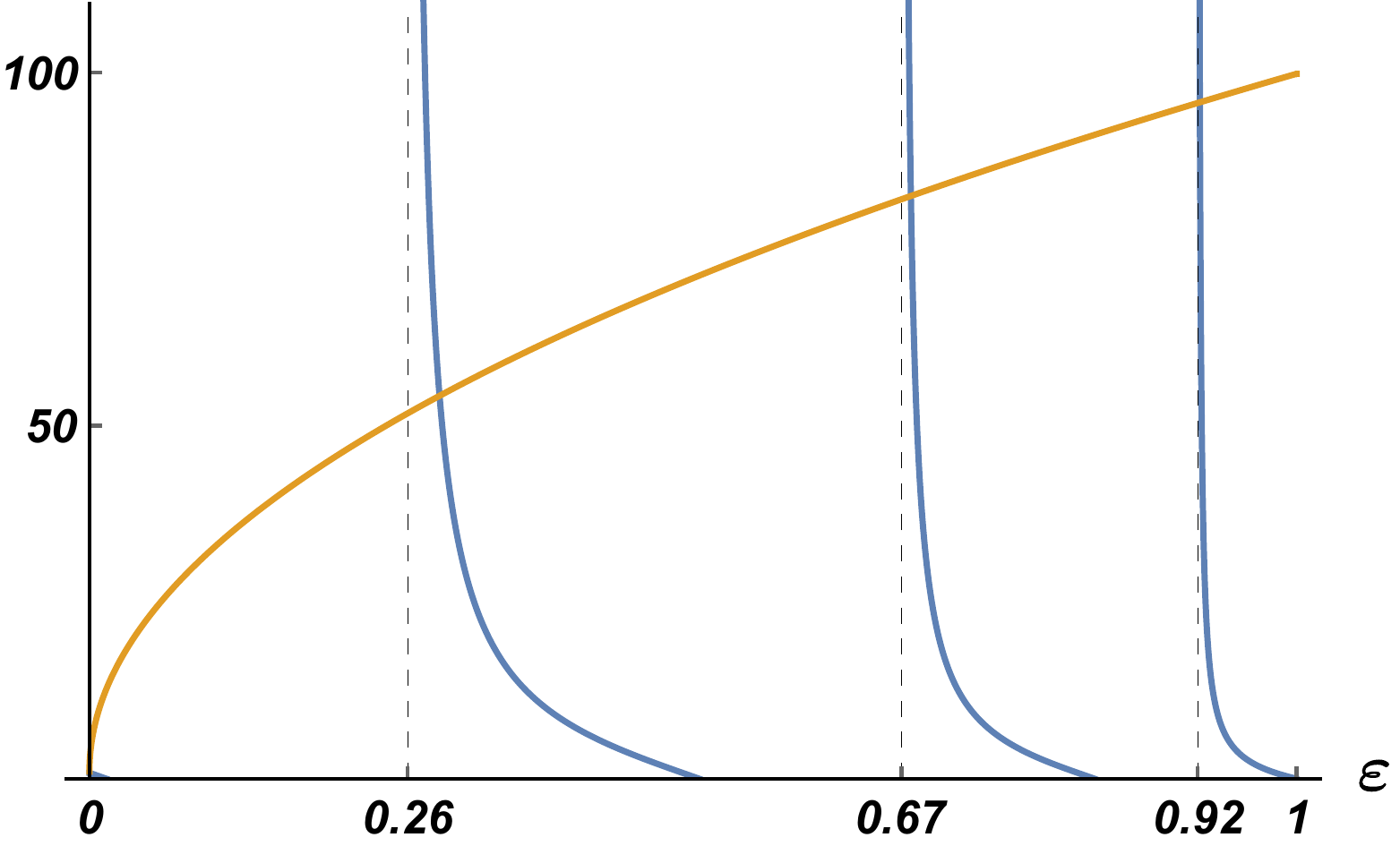}
}
\qquad
\resizebox{0.3\textwidth}{!}{
 \includegraphics{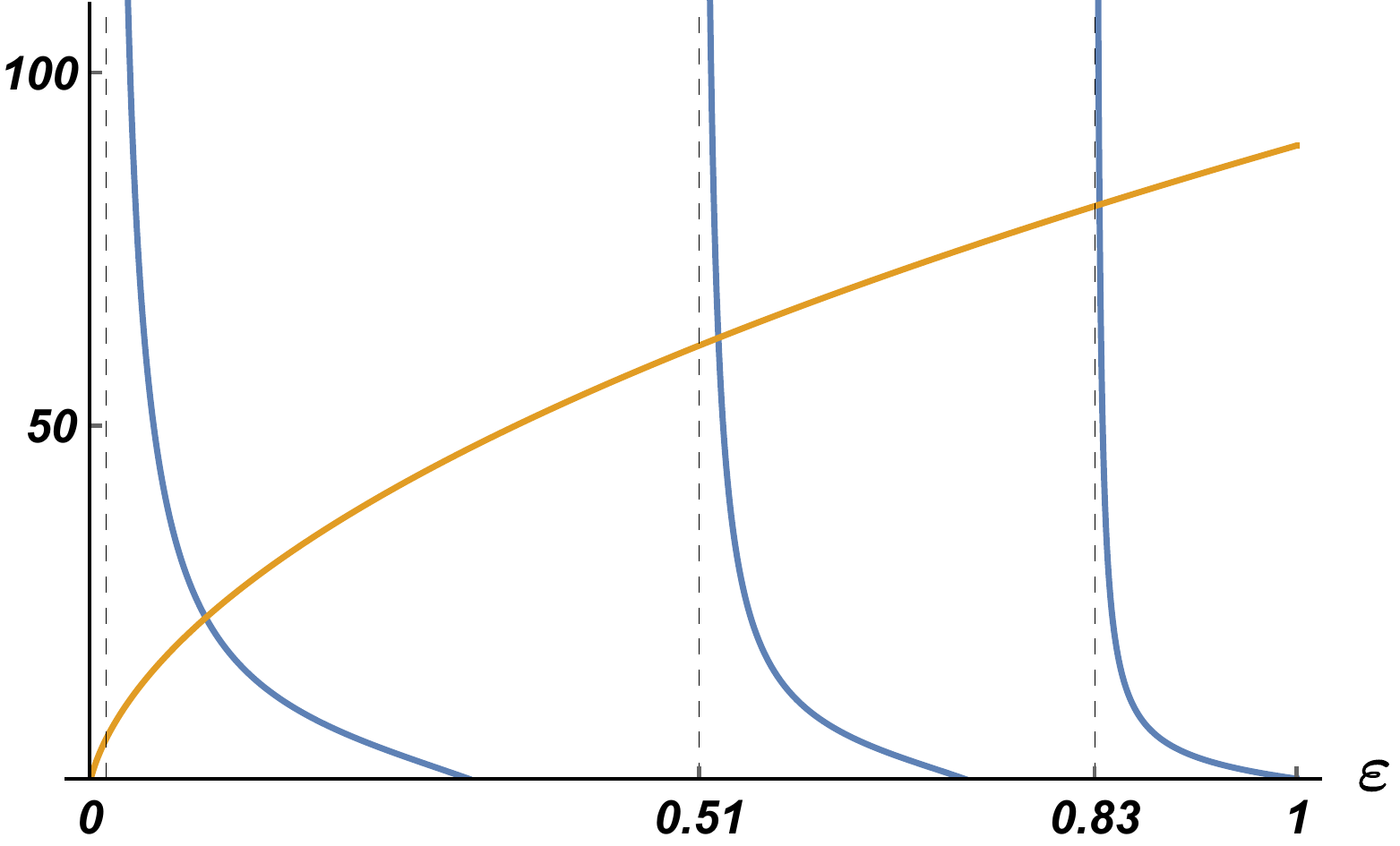}
}
\caption{\small 
Results for the isotope ${}^{209}$Pb. In blue the left-hand side $\varphi_\ell$ of \eqref{2.16} in yellow the right-hand side $\phi_\ell$ for $\beta=1$. On the left the case $\ell=0$, on the right the case $\ell=1$ and {$j=3/2$}.}
\label{fig_Pb}       
\end{figure}

Now we add the $\delta'$ interaction in the nucleus ${}^{209}$Pb.
In Figure~\ref{fig_Pb}, we use the  parameters given above  for $\ell=0$ and $\ell=1$, where the energy of the bound states is given in Table \ref{table.Pb}. We also compare these results with those obtained  with the Gamow code. We have chosen $\beta=0$ and $\beta=1$, which corresponds to $V_\text{q}=0$ and $V_\text{q}=-20.83$~MeV~fm$^2$, respectively. The numerical approximation of the square well and singular potentials in Table~\ref{table.Pb} are simulated taking $a=0.01$~fm.
From Table~\ref{table.Pb}, we observe that there are three bound states for  both values $\beta=0$ and $\beta=1$. The same number of bound states is obtained when we consider the $\delta$-$\delta'$ M, see Figure~\ref{fig_Pb}. 
In Table~\ref{table.Pb} we have observed some discrepancies between the results obtained with our formalism and the numerical calculations obtained when considering the $\delta'$ interaction as a limit of odd functions in the Gamow code. This is to be expected and the origin lies in the different definitions of the $\delta'$ interaction explained at the end of section \ref{sec.mf}. Nevertheless, our intention is to show how these differences vary with the  quantum numbers $(n,\ell,j)$.
Indeed, in the data of Table~\ref{table.Pb} we can  verify that the inequalities of \eqref{eq:pro3} are always satisfied when the $\delta'$ term is added.

\begin{table}[h!]
	\centering
	\caption{Comparison of the neutron energy levels (MeV) in the core of $^{208}$Pb between the numerical Gamow code results \cite{1982Vertse} versus our model ($\delta$-$\delta'$ M) results, using the physical parameters mentioned in the text. Calculations are done for $\beta=0$ (second and third columns) and $\beta=1$ (fourth and fifth columns).}
	\label{table.Pb}       
	\begin{tabular}{lllll}
		\hline\noalign{\smallskip}
		\multirow{2}[3]{*}{State} & \multicolumn{2}{c}{$\beta=0$} & \multicolumn{2}{c}{$\beta=1$} \\
		\cmidrule(lr){2-3} \cmidrule(lr){4-5}
		& Gamow &  $\delta$-$\delta'$ M   & Gamow & $\delta$-$\delta'$ M \\
		\noalign{\smallskip}\hline\noalign{\smallskip}
		$0s_{1/2}$   & -41.35    &  -41.36   & -40.97   &  -40.84 \\
		$1s_{1/2}$   & -32.27    &  -32.31   & -31.11   &  -30.23  \\
		$2s_{1/2}$   & -17.53    &  -17.61   & -18.11   &  -12.91  \\[0.25ex]
		\hdashline\\[-2.25ex]
		$0p_{1/2}$  & -38.08    &   -37.80  & -37.34   &  -37.10 \\
		$1p_{1/2}$   & -25.91    &  -25.04   & -24.44   &  -22.90 \\
		$2p_{1/2}$   & -8.47   & -6.67  &  -11.20  & -2.38 \\[0.25ex]
		\hdashline\\[-2.25ex]
		$0p_{3/2}$    & -38.21    &   -38.54  & -37.48   &   -37.15 \\
		$1p_{3/2}$    & -26.29    &   -27.18  & -25.30   &   -23.16  \\
		$2p_{3/2}$    & -9.17  &   -10.63  & -13.30   &   -4.31  \\
		\noalign{\smallskip}\hline
	\end{tabular}
\end{table}

\section{Concluding remarks} 

We have studied a spherical well plus a linear combination of a Dirac delta and a local $\delta'$ interaction, both located at the well edge. 
Due to spherical symmetry, the problem   reduces to a one-dimensional one, by means of the radial Schr\"odinger equation. This contact potential has been defined by using appropriate matching conditions satisfied by the radial wave functions. 
In particular, we have obtained general and precise properties concerning the number and behaviour of bound states.  These  are summarized in the three theorems of section \ref{section4}. Note that due to the singular character of the studied interaction, general results applicable to well-behaved spherical potentials can not be used. Nevertheless, we have been able to extend some of them to our case. In addition, we have even obtained some precise analytical expressions, as for instance, we not only guarantee the existence of $\ell_{\rm max}$, we provide a specific expression. We have also presented the simplifications in the bound states structure for certain configurations of the parameters of the model. 

We have found that the dependence of bound states with the coefficient $\alpha$ of the $\delta$ potential is stronger than the dependence on the coefficient $\beta$  of the $\delta'$ interaction. There are, nevertheless, some exceptions, the most interesting occurs when $\beta$ reaches two critical points, yielding to the appearance of Robin or Dirichlet BC.

This model has also antibound states and resonances. 
These are characterized by the existence of poles in the analytic continuation, $S(k)$, of the $S$-matrix in the moment representation. These poles may also be obtained using the so-called {\it purely outgoing boundary conditions}, which are determined by equating to zero the coefficient of the asymptotic form of the incoming wave. This coefficient depends on the momentum $k$, which gives a transcendental equation, for which the solutions are the poles of $S(k)$.
Exact and numerical values for resonances, bound and antibound states can be obtained for all values of the orbital angular momentum, although the case $\ell=0$ is by far the simplest.

In the last section we have used this configuration to approximately describe the extra neutron energy levels of a double magic nucleus with  spin-orbit interaction  plus an extra mean-field interaction, testing the numerical results with the nuclei  $^{209}$Pb and $^{133}$Sn. We have shown that the Hamiltonian \eqref{eq.sh}, inspired by the Woods-Saxon potential after the limit $a\to 0^+$, gives a good approximation for the low-lying bound states. 
The $\delta$ term  gives the nuclear spin-orbit contribution. The aim of the additional interaction,   given by the $\delta'$ interaction, is providing a correction such that the results of the proposed model better fit to the experimental data, in particular for states like the ones shown in Table~\ref{table.eexp}.
This simplified model could be used to gain insight into the neutron energy levels since the main advantage over the Woods-Saxon potential is that we can solve exactly the eigenfunction equation, obtaining analytic properties of the spectrum using well-known features of  Bessel functions. In any case, our goal is to describe  properties in a qualitative manner,  we do not expect to get a numerical fit with good precision.

Along our discussion, we have  mentioned that there are two possible choices of the $\delta'$ interaction. As pointed out before, we have chosen the only one which is compatible, the resulting Hamiltonian is self adjoint, with the $\delta$ interaction supported at the same point, the so called local $\delta'$ interaction.  We have obtained some numerical results, which slightly deviate from those obtained using the regular mean-field potential. 
The reason is that the limit $a\to 0^+$ in \eqref{eq.vdp}  leads to a $\delta'$ potential which  does not give a self adjoint version of the Hamiltonian.

As a final remark, we may explore similar approximations with potentials of another type which may be also interesting in nuclear systems, in the nearest future. One possibility is to replace the three dimensional Wood-Saxon potential by the three dimensional Scarf II potential as studied by L\'evai and coworkers \cite{LEV}.

\section*{Acknowledgements}

This work was  supported by Consejo Nacional de Investigaciones Cient\'{\i}ficas y T\'ecnicas PIP-625 (CONICET, Argentina), the Spanish MINECO (MTM2014-57129-C2-1-P), Junta de Castilla y Le\'on and FEDER projects (BU229\-P18 and VA137G18). C.R. is grateful to MINECO for the FPU fellowships programme (FPU17/01475).

\begin{appendices}
	
\numberwithin{equation}{section}
\section{On the self adjointness of the Hamiltonian}\label{A}

The goal of the present appendix is a discussion on the self adjointness of the radial Hamiltonian \eqref{2.2}. Setting the appropriate units such that $\hbar=1$ and $2 \mu=1$ this Hamiltonian may be written as
\begin{equation}\label{8.1}
H=-\frac{d^2}{dr^2}+ [\theta(r-R)-1]V_0+a\delta(r-R) +b\delta'(r-R)+\frac{\ell(\ell+1)}{r^2}\,,
\end{equation}
$\ell\in\mathbb{N}_0$. Let us split it into $H=H_\ell+V(r)$, where
\begin{equation}\label{8.2}
H_\ell=-\frac{d^2}{dr^2}+\frac{\ell(\ell+1)}{r^2}.
\end{equation}
 For  the sake of clarity, we first study $H_{\ell=0},$ which reduces to the one-dimensional Laplace operator in a given domain.
 
\subsection{Zero angular momentum}

We have to find a domain for $H_0$, which must be a subspace of $L^2[0,\infty)$. This domain must include all square integrable absolutely continuous functions, $f(r)$, with absolutely continuous derivative and square integrable second derivative. Thus,
\begin{equation}\label{8.3}
\int_0^\infty \{|f(r)|^2+|f''(r)|^2\}\,dr<\infty\,.
\end{equation}
The boundary conditions at the origin should be specified in such a way that $H_0$ is Hermitian on its domain. In consequence, for any $f(r),g(r)$ in the domain of $H_0$, 
\begin{eqnarray}
\langle h(r) | H_0 \, f(r) \rangle 
      			&=& -\int_0^\infty h^*(r)\,f''(r)\,dr 
			=h^*(0)\,f'(0) - {h'}^*(0)\,f(0)
      						- \int_0^\infty {h''}^*(r)\,f(r)\,dr  \nonumber \\
&=& h^*(0)\, f'(0) - {h'}^*(0)\, f(0) 
              			+\langle H_0 \, h(r) |f(r) \rangle.
\label{8.4}
\end{eqnarray}
Then, $H_0$ is Hermitian in the given domain if, and only if, $h^*(0)\,f'(0)-{h'}^*(0)\,f(0)=0$, which happens if, and only if, $f(0)=cf'(0)$ for any function $f(r)$ in this domain, where $c$ is an arbitrary real constant. For $c=0$, we have that $f(0)=0$ with $f'(0)$ arbitrary. Since $c^{-1}f(0)=f'(0)$, another possible choice is $f'(0)=0$ with $f(0)$ arbitrary. Here, we may say that $c=\infty$. All these possible choices select a domain, $\mathcal D$, in which  $H_0$ is self adjoint. We select any one of them. 

After selecting a value of $c\in \mathbb R\cup\{\infty\}$, let us consider a subspace of $\mathcal D$, denoted by $\mathcal D(H_0)$. By definition, $f(r) \in \mathcal D(H_0)$ if, and only if,, $f(R)=f'(R)=0$. Choosing $\mathcal D(H_0)$ as the domain of $H_0$, we see that $H_0$ is symmetric (Hermitian), although not self adjoint, having deficiency indices $(2,2)$. 

In order to prove this latter statement, let us recall that the domain of the adjoint $H_0^\dagger$  is determined by
\begin{eqnarray}
{\mathcal D}(H_0^\dagger)= \{ h(r)\in L^2[0,\infty);\; \exists\,g(r)\in L^2[0,\infty); 
\langle h(r)|H_0\,f(r)\rangle=\langle g(r)|f(r)\rangle\},  \label{8.5}
\end{eqnarray}
for all $f(r)$ in $\mathcal D(H_0)$. To obtain a basis of the deficiency subspaces \cite{RSII}, we have to solve the equations $h''(r)=\pm ih(r)$, where the solutions must be in $\mathcal D(H_0^\dagger)$. Let us choose the sign plus first. We obtain two linearly independent solutions, which are:
\begin{eqnarray*}\label{8.6}
h_1(r)=\left\{ \begin{array}{ccc} C\,e^{-\frac{\sqrt 2}2\,r} \,e^{-i\frac{\sqrt 2}2\,r}, &  & r>R, \\[1ex]  
0, &  & r<R,  \end{array}   \right.
\qquad h_2(r)= \left\{ \begin{array}{ccc}    0, &  &  r>R,\\[1ex]  
A\,   e^{-\frac{\sqrt 2}2\,r} \,e^{-i\frac{\sqrt 2}2\,r}+ \frac{A(1-c)}{1+c}\, e^{\frac{\sqrt 2}2\,r} \,e^{i\frac{\sqrt 2}2\,r},  &  & r<R,        
\end{array}   \right.
\end{eqnarray*}
where $A$ and $C$ are arbitrary constants. The linear independence of these two functions is obvious, so that they are a basis for the deficiency subspace corresponding to the plus sign. Similar analysis can be performed for the minus sign. This proves that the deficiency indices for $H_0$ with domain $\mathcal D(H_0)$ are precisely $(2,2)$. In this circumstance, $H_0$ admits an infinite number of self adjoint extensions labelled by four independent real parameters. Domains for these self adjoint extensions are determined by matching conditions at the point $r=R$ as usual \cite{KU}, where the exceptional cases $\beta=\pm 2$ are also included. The choice of the matching conditions \eqref{2.15} gives a two parametric family of self adjoint extensions, which proves the self adjointness of 
\begin{equation}\label{8.8}
H_r=-\frac{d^2}{dr^2}+ a\delta(r-R) +b\delta'(r-R)\,,
\end{equation}
which is \eqref{8.1} with $\ell=0$ and without the term  $V_0\,[\theta(r-R)-1]$. As we will explain at the end of the present appendix, adding this term to the potential does not change the self adjointness.

\subsection{Higher angular momentum}

For $\ell\geq 1$ we do not need to impose conditions at the origin of the type $f(0)=cf'(0)$, as the Hamiltonian \eqref{8.2} is essentially self adjoint when its domain is given by the Schwartz space supported on $\mathbb R^+:=[0,\infty)$. For these functions $f(0)=f'(0)=0$, so that  $h^*(0)\,f'(0)-{h'}^*(0)\,f(0)$ is automatically zero. Then, we define $\mathcal D(H_\ell)$, $\ell\ne 0$, to be the space of functions $f(r)\in L^2[0,\infty)$ satisfying the following conditions \cite{AJS}:
\begin{enumerate}
\item $f(r)$ and $f'(r)$ are absolutely continuous.
\item $-f''(r)+[\ell(\ell+1)/r^2]f(r)$ is square integrable, \textit{i.e.}, it belongs to $L^2[0,\infty)$. 
\item $f(0)=0$. 
\item $f(R)=f'(R)=0$. 
\end{enumerate}

In order to obtain the deficiency subspaces for $H_\ell$, we have to find the square integrable solutions of the following pair of differential equations:
\begin{equation}\label{8.9}
h''(r)-\frac{\ell(\ell+1)}{r^2}\,h(r)\mp ih(r)=0\,.
\end{equation}

For the minus sign in \eqref{8.9}, the general solution is given by \cite{AJS} (p. 478):
\begin{equation}\label{8.10}
u(r)=A\,r^{1/2}\,J_{\ell+1/2}(r\sqrt i)+B\,r^{1/2}\, Y_{\ell+1/2}(r\sqrt i)\,,
\end{equation}
where $J_{\ell+1/2}$ and $Y_{\ell+1/2}$ are the Bessel and Neumann functions \cite{AS}, respectively. Asymptotic properties of these functions show that \cite{AS,AJS}:
\begin{eqnarray*}
r^{1/2}\,J_{\ell+1/2}(r\sqrt i)\notin L^2[1,\infty)\,, &\quad & r^{1/2}\,J_{\ell+1/2}(r\sqrt i) \in L^2[0,1], \label{8.11}\\[1ex]
r^{1/2}\, Y_{\ell+1/2}(r\sqrt i) \in L^2[1,\infty)\,, &\quad & r^{1/2}\, Y_{\ell+1/2}(r\sqrt i) \notin L^2[0,1] \label{8.12}\,.
\end{eqnarray*}

Therefore, the basis for the deficiency subspace with minus sign in \eqref{8.9} is given by the following pair of functions:
\begin{eqnarray*}\label{8.13}
u_1(r)=\left\{ \begin{array}{ccc}     
A\,  r^{1/2}\, Y_{\ell+1/2}(r\sqrt i), & {\rm if}  &  r>R,  \\[1ex]  
0, & {\rm if}  & r<R ,           
\end{array}        \right.
\qquad u_2(r)=\left\{ \begin{array}{ccc}     0,  & {\rm if }  & r>R, \\[1ex]  
B\,r^{1/2}\,J_{\ell+1/2}(r\sqrt i), & {\rm if}  & r<R  ,        
\end{array}      \right.
\end{eqnarray*}
where $A$ and $B$ are constants. A similar result can be obtained for the plus sign in \eqref{8.9}, so that the deficiency indices for $H_\ell$ with $\ell\ne 0$ are $(2,2)$. Self adjoint extensions are obtained by suitable matching conditions at $r=R$ and depend on four real parameters. Again, the choice of matching conditions \eqref{2.15}, where the exceptional cases $\beta=\pm 2$ are included, determines a self adjoint Hamiltonian of the form,
\begin{equation}\label{8.15}
H_r:=-\frac{d^2}{dr^2}  +a\delta(r-R) +b\delta'(r-R)+\frac{\ell(\ell+1)}{r^2}\,,\quad \ell\ne 0\,,
\end{equation}
which is  \eqref{8.1} without the term $V_0\, [\theta(r-R)-1]$. Adding this term does not change anything in both cases ($\ell=0$ and $\ell\ne 0$). Once we have determined the domains for which \eqref{8.8} and \eqref{8.15} are self adjoint, since the term $V_0\, [\theta(r-R)-1]$ is bounded and Hermitian, it is self adjoint. Now, the Kato-Rellich theorem \cite{RSII} says that if $H_r$ is self adjoint, so is $H_r+[\theta(r-R)-1]V_0$. We conclude that it is possible to determine domains such that \eqref{8.1} is self adjoint for all values $\ell\in\mathbb{N}_0$.

\section{Proofs of theorems 1, 2 and 3}

\subsection{Proof of theorem 1}\label{B4}

In the first place, we show that the right-hand side of \eqref{2.16} is positive and strictly growing as a function of the relative energy. As will be proved in theorem \ref{prop:5},  there exists an upper bound for the angular momentum, $\ell_\text{max}$ hence the term $8\beta(\ell+1)$ is always finite.   From Theorem~6 in \cite{SEG} there exits the following bounds for the following ratio of  modified Bessel functions:
\begin{equation}\label{4.5}
\sqrt{\sigma^2+\ell^2} +\ell +1 \le \frac{\sigma\,K_{\ell+3/2}(\sigma)}{K_{\ell+1/2}(\sigma)}< \sqrt{\sigma^2+(\ell+1)^2}+\ell+1.
\end{equation} 
Now we can use the first inequality of \eqref{4.5}  together with \eqref{eq:Boundw_0} to derive:
\begin{eqnarray*}
\phi_\ell(\sigma)= \frac{(2+\beta)^2}{(2-\beta)^2}\, \frac{\sigma\,K_{\ell+3/2}(\sigma)}{K_{\ell+1/2}(\sigma)} - \frac{8\beta(\ell+1)}{(2-\beta)^2}+\frac{w_{\ell j}}{(2-\beta)^2}  > 
\frac{(\beta -2)^2+2 \ell\left(\beta ^2+4\right) +w_{\ell j}}{(\beta -2)^2}  >0.
\end{eqnarray*}
In addition, using the Turan-type inequalities given in \cite{LANATA}, we can prove the following relation:
\begin{equation}\label{7.14}
\frac{d \phi_\ell(\sigma)}{d\sigma}= 
\frac{\sigma\,K_{\ell-1/2}(\sigma) \,K_{\ell+3/2}(\sigma) -\sigma\, K^2_{\ell+1/2}(\sigma)}{K^2_{\ell+1/2}(\sigma)} >0\,.
\end{equation}
This shows that $\phi_\ell(\sigma)$ is a strictly growing positive function on the variable $\sigma$ and, due to the definition of $\sigma(\varepsilon)$ \eqref{3.1}, on $\varepsilon$.

On the other hand, if we show that $\varphi_\ell(\chi)$, the left-hand side of \eqref{2.16}, is one to one and onto as a function between the following intervals:
\begin{equation}\label{7.15}
\varphi_\ell(\chi) : (j_{\ell+3/2,s-1}, j_{\ell+1/2,s})\subset (0,v_0) \mapsto (0,\infty), \  s\in \mathbb{N},
\end{equation}
we guarantee the unique existence of the bound state $\varepsilon_s$ in \eqref{3.8}.
Thus, it will be enough to demonstrate that $\varphi_\ell(\chi)$ is strictly monotonic on $\chi$ and that it covers the whole $(0,\infty)$ as $\chi \in (j_{\ell+3/2,s-1}, j_{\ell+1/2,s})$. In fact, the first derivative of $\varphi_\ell(\chi)$ meets
\begin{equation*}\label{7.16}
\frac{d\varphi_\ell(\chi)  }{d\chi}\,
= \chi\; \frac{J^2_{\ell+1/2}(\chi) - J_{\ell-1/2}(\chi)\,J_{\ell+3/2}(\chi)}{J^2_{\ell+1/2}(\chi)}>0\,,
\end{equation*}
where the equality follows from  standard properties of the Bessel functions \cite{OLVER} and the second relation from the Turan-type inequalities \cite{LANATA,MUNIRO}:
\begin{equation}\label{7.17}
J^2_n(\chi)-J_{n-1}(\chi)\,J_{n+1}(\chi) > \frac{J^2_n(\chi)}{n+1}>0\,,
\end{equation}
where $\chi$ is real and $n>0$, with $n=\ell+1/2$. Finally, in the given intervals,  the function $\varphi_\ell(\chi)$ is positive and
\begin{eqnarray}\label{7.18}
\lim_{\chi\to j_{\ell+3/2,s-1}} \varphi_\ell(\chi)=\varphi_\ell(j_{\ell+3/2,s-1})=0\,,\qquad \lim_{\chi\to j_{\ell+1/2,s}^-} \varphi_\ell(\chi) =\infty\,.
\end{eqnarray}

Now we focus on the second part of the theorem concerning the number of bound states \eqref{Prop4nl}. The key feature is the existence of an integer $s_0$ for which  
$$(j_{\ell+3/2,s-1}, j_{\ell+1/2,s}) \cap (0,v_0)=\emptyset, \ \forall s\geq s_0.
$$
Let us examine this in greater detail. The largest integer $M$ for which $(j_{\ell+3/2,M-1}, j_{\ell+1/2,M})\subset (0,v_0)$ still holds is obviously given by
\begin{equation}
j_{\ell+1/2,M}< v_0,\qquad j_{\ell+1/2,M+1}> v_0.
\end{equation}

Since $\varphi_\ell(\chi(\varepsilon))$ is strictly decreasing and $\phi_\ell(\sigma(\varepsilon))$ strictly increasing as functions of $\varepsilon$, 
the condition
\begin{eqnarray}\label{eq:limit0Prop3}
\varphi_\ell(\chi(\varepsilon))=\varphi_\ell(v_0)={v_0  J_{\ell+\frac{3}{2}}(v_0 )}/{J_{\ell+\frac{1}{2}}(v_0 )}  >\dfrac{(\beta -2)^2+2 \ell\left(\beta ^2+4\right) +w_{\ell j}}{(\beta -2)^2}=\underset{\varepsilon \to 0^+}{\lim}\phi_\ell(\sigma(\varepsilon)).  \nonumber 
\end{eqnarray}
implies the existence of an additional bound state whose energy is the closest to $\varepsilon=0$. In the particular case  $v_0=j_{\ell+1/2,M}$  no additional bound state should be added to $M$.
Independently, if
\begin{equation}\label{eq:limit1Prop3}
\underset{\varepsilon \to 1^-}{\lim}\varphi_\ell(\chi(\varepsilon))=0
>\underset{\varepsilon \to 1^-}{\lim}\phi_\ell(\sigma(\varepsilon))
\end{equation}
no bound state satisfying $\varepsilon\in (1-j^2_{\ell+1/2,1}/v_0^2, 1)$ appears. Only if $j_{\ell+1/2,1}>v_0$ the functions $m$ and $m'$ defined in the present theorem are not independent. Nevertheless, a bound state with relative energy $\varepsilon\in(0,1)$ appears if, and only if, $\varphi_\ell(v_0)>\underset{\varepsilon \to 0^+}{\lim}\phi_\ell(\sigma)$ and $0
<\underset{\varepsilon \to 1^-}{\lim}\phi_\ell(\sigma)$ so that $n_\ell$ is also given by \eqref{Prop4nl} and the proof is concluded.

\subsection{Proof of theorem \ref{prop:5}}\label{B5}

For this proof we use  the number of bound states given in equation \eqref{Prop4nl} of theorem~\ref{prop:4} in order to obtain
\begin{equation}
n_\ell=0 \qquad \forall \ \ell>\ell_{\max}.
\end{equation}
Note that in the derivation of \eqref{Prop4nl}, we have not used the assumption of the existence of $\ell_\text{max}$ that appears at the beginning of appendix~\ref{B4} so there is   no circular reasoning. Due 
to the properties of the zeros of the Bessel function \eqref{3.7} and their asymptotic expressions for large order \cite{OLVER}, there exists an integer $\ell_0$ such that
\begin{equation}\label{eq.98}
	j_{\ell+1/2,1}>v_0\, , \qquad \ell\geq \ell_0,
\end{equation}
and therefore $M=0$. Eventually, we shall reach a value $\ell_\text{max}\geq \ell_0$ such that
\begin{eqnarray}
\underset{\varepsilon \to 0^+}{\lim}\varphi_\ell(\chi(\varepsilon))=\varphi_\ell(v_0)={v_0  J_{\ell+\frac{3}{2}}(v_0 )}/{J_{\ell+\frac{1}{2}}(v_0 )}   <\frac{(\beta -2)^2+2 \ell\left(\beta ^2+4\right) +w_{\ell j}}{(\beta -2)^2}=\underset{\varepsilon \to 0^+}{\lim}\phi_\ell(\sigma(\varepsilon)), \nonumber
\end{eqnarray}
for all $\ell>\ell_{\max}$, hence $m=m'=0$. In effect, the existence of $\ell_\text{max}$ is a consequence of the dependence on the angular momentum of both sides in the previous inequality. It is clear that the right-hand side is a strictly increasing function with respect to $\ell$. In addition, using Theorem~3 of  \cite{LAN} it can be easily proved that the left-hand side fulfils 
\begin{equation}\label{eq:th3}
 \dfrac{\partial \varphi_\ell(v_0)}{\partial \ell} \leq 0.
\end{equation}
In consequence,  if none of the conditions $j_{\ell+1/2,1}<   v_0$, $\varphi_\ell(v_0)>\phi_\ell(0^+)$ hold, there is no bound state.

In order to complete the proof, we should consider a configuration in which  an integer $s_0 \in \mathbb{N}$ such that $v_0= j_{\ell+1/2,s_0}$ exists. In such a case  the condition $\varphi_\ell(v_0)>\phi_\ell(0^+)$ is ill defined. Nevertheless, if $s_0>1$ we know the existence of at least   one bound state. Thus, we have to consider the next value of the angular momentum for which $v_0\neq j_{\ell+1+1/2,s_0}$. 
 If $s_0=1$ the bound state  always exists, although, if $\phi_\ell(v_0)<0$, its energy can be below $-V_0$. Thus,  we have to consider the next value of the angular momentum for which $v_0\neq j_{\ell+1+1/2,1}$.

\subsection{Proof of theorem \ref{prop:6}}\label{B6}

Throughout  the proof we bear in mind the   monotonicity properties of $\varphi_\ell(\varepsilon)$ and $\phi_\ell(\varepsilon)$ with respect to $\varepsilon$  demonstrated in appendix~\ref{B4}. 
\begin{itemize}
\item[$(a)$] 
The inequality $(a)$ in \eqref{eq:pro3} is just a consequence of $j_{\lambda,i}<j_{\lambda,i+1}$ given in \eqref{3.7}. 

\item[$(b)$] 
To prove the inequality $(b)$ in \eqref{eq:pro3} we first take into account that the bound states characterized by   $n$ are determined, for a given $\ell$,  by the function $\varphi_\ell(\varepsilon)$ restricted to the interval $(a_{n\ell},b_{n\ell})$, where
$$
a_{n\ell}=1- \frac{j^2_{\ell+1/2,n+1}}{v_0^2}  , 
\quad
b_{n\ell}=1- \frac{j^2_{\ell+1/2,n}}{v_0^2}  , \quad  n \in \mathbb{N}_0.
$$
We need to consider both functions $ \varphi_\ell (\varepsilon)$, $ \varphi_{\ell+1} (\varepsilon)$, and therefore both intervals $(a_{n\ell},b_{n\ell})$,  and $(a_{n(\ell+1)},b_{n(\ell+1)})$. Due to the properties of the zeros $ j_{\ell+1/2,n }$ given in \eqref{3.7}, the following relations are fulfilled:
\begin{equation*}
a_{n(\ell+1)}< a_{n\ell}, \qquad b_{n(\ell+1)}< b_{n\ell}.
\end{equation*}
Therefore, either (i) $b_{n(\ell+1)}\leq a_{n\ell}$, or (ii) $a_{n\ell}<b_{n(\ell+1)}$. 

\begin{itemize}
\item[$\bullet$]
If (i) is true, 
$(a_{n(\ell+1)},b_{n(\ell+1)}) \cap (a_{n\ell},b_{n\ell})=\emptyset $ so $-\varepsilon_{n\ell_{j}}<-\varepsilon_{n(\ell+1)_{j}}$ holds trivially.
\item[$\bullet$]
If (ii) is true, then we have three disjoint intervals: $(a_{n(\ell+1)},a_{n\ell})$, $(a_{n\ell},b_{n(\ell+1)})$ and $(b_{n(\ell+1)},b_{n\ell})$. If
$\varepsilon_{n\ell_{j}}\in (b_{n(\ell+1)},b_{n\ell})$ or
$\varepsilon_{n(\ell+1)_{j}}\in (a_{n(\ell+1)},a_{n\ell})$, then it is obvious that $-\varepsilon_{n\ell_{j}}<-\varepsilon_{n(\ell+1)_{j}}$. However, if  
$\varepsilon_{n\ell_{j}}, \varepsilon_{n(\ell+1)_{j}}\in (a_{n\ell},b_{n(\ell+1)})$, the situation needs to be studied in detail.
Let us prove first:
\begin{equation}\label{eq:ineqs}
\varphi_{\ell}(\varepsilon_0)\geq\varphi_{\ell+1}(\varepsilon_0),\quad 
\phi_{\ell}(\varepsilon_0)<\phi_{\ell+1}(\varepsilon_0), 
\end{equation}
$\varepsilon_0\in(a_{n\ell},b_{n(\ell+1)})$. The first part results from \eqref{eq:th3}, since it holds for  $v_0\in \mathbb{R}$, excluding the singularities of $\varphi_\ell$ \cite{LAN}. 
For the second case,  the bounds  in \eqref{4.5} ensure
\begin{eqnarray*}
&& \frac{\sigma\,K_{(\ell+1)+3/2}(\sigma)}{K_{(\ell+1)+1/2}(\sigma)}-\frac{\sigma\,K_{\ell+3/2}(\sigma)}{K_{\ell+1/2}(\sigma)}   > 1.
\end{eqnarray*}
Consequently, using \eqref{2.16}, we reach
\begin{equation}\label{eq:w0eqs0}
\phi_{\ell+1}(\varepsilon_0) - \phi_{\ell}(\varepsilon_0) > 1+\frac{w_{(\ell +1) j}-w_{\ell j}}{(2-\beta)^2}.
\end{equation}
In addition, for $\ell +1$, $j=(\ell+1)-1/2$ so, bearing in mind \eqref{xiellj}, the parameter $w_{(\ell +1) j}>0$. In a similar way, for $\ell$, $j=\ell+1/2$ and $w_{\ell j}<0$. Consequently, the second inequality in \eqref{eq:ineqs} is proved. 

Now, we may prove $\varepsilon_{n(\ell+1)_j} < \varepsilon_{n\ell_j} $ by contradiction. Let us assume $\varepsilon_{n(\ell+1)_j} \geq \varepsilon_{n\ell_j}$.  With \eqref{eq:ineqs} and the monotonicity with respect to $\varepsilon$ above-men\-tioned, we find
\begin{eqnarray}
\phi_\ell(\varepsilon_{n\ell_j})\leq\phi_\ell(\varepsilon_{n(\ell+1)_j}) < \phi_{\ell+1}(\varepsilon_{n(\ell+1)_j}) 
 =\varphi_{\ell+1}(\varepsilon_{n(\ell+1)_j}) &\leq &\varphi_{\ell+1}(\varepsilon_{n\ell_j})\leq\varphi_{\ell}(\varepsilon_{n\ell_j}). \nonumber 
\end{eqnarray}
From here it follows  $\varphi_\ell(\varepsilon_{n\ell_j})\neq \phi_{\ell}(\varepsilon_{n\ell_j})$,
which is clearly absurd because $\varepsilon_{n\ell_j}$ is a bound state, and the equality, \eqref{2.16}, must be satisfied.

\end{itemize}

\item[$(c)$] 
The inequality $(c)$ in \eqref{eq:pro3} is  proved taking into account that the only dependence on $j$ in the secular equation \eqref{2.16} is through $w_{\ell j}$. As we have already pointed out, $w_{\ell(\ell-1/2)}>0 >w_{\ell(\ell+1/2)}$. In consequence, $\phi_\ell$ is greater for $j=\ell-1/2$. Since $\varphi_\ell$ is independent of $j$ we only need to consider the interval $(a_{n\ell},b_{n\ell})$, for which  the inequality is proved, as has been done before, by contradiction.

\end{itemize}

\end{appendices}

\end{document}